\documentclass[12pt,a4paper,psfig]{article}
\input{psfig.sty}
\usepackage[dvips]{graphicx}

\begin{document}

\title{Cosmic  gamma-ray bursts: observations and modeling.}
\author{G.S.Bisnovatyi-Kogan
\thanks{Institute for Space Research, Russian Academy of Sciences;
and Joint Institute for Nuclear Research, Dubna; Email:
gkogan@iki.rssi.ru}}

\date{}
\maketitle
\begin{abstract}

It is now commonly accepted that cosmic $\gamma$ -ray bursts (GRBs)
are of cosmological origin. This conclusion is based on the
statistical analysis of GRBs and the measurements of line redshifts
in GRB optical afterglows, i.e., in the so-called long GRBs. In this
review, the models of radiation and models of GRB sources are
considered. In most of these models, if not in all of them, the
isotropic radiation cannot provide the energy release necessary for
the appearance of a cosmological GRB. No correlation is noted
between the redshift, the GRB-spectrum shape, and the total detected
energy. The comparison between data obtained in the Soviet
experiment KONUS and the American experiment BATSE shows that they
substantially differ in statistical properties and the detection of
hard x-ray lines. The investigation of hard gamma (0.1–10 GeV)
afterglows, the measurement of prompt optical spectra during the GRB
detection, and the further investigation of hard x-ray lines is of
obvious importance for gaining insight into the GRB origin.
Observations of two bright optical GRB afterglows point to the fact
that an initially bright optical flare is directly related to the
GRB itself, and the subsequent weak and much more continuous optical
radiation is of a different nature. The results of observations of
optical GRB afterglows are discussed. They point to the fact that
the GRBs originate in distant galaxies with a high matter density,
where intense star formation takes place. The interaction of the
cosmological GRB radiation with a dense surrounding molecular cloud
results in the appearance of long-duration (up to 10 years) weak
optical afterglows associated with the heating and reradiation of
gas. Results of 2D numerical simulation of the heating and
reradiation of gas in various variants of the relative disposition
of GRB and molecular clouds are presented. In conclusion, the
possible relation between the so-called short GRBs and recurrent
sources of soft $\gamma$ rays in our Galaxy, the so-called “soft
gamma repeaters,” is discussed.

\end{abstract}

\section{Introduction}

Nowadays, it is commonly accepted that cosmic $\gamma$ -ray bursts
(GRBs), the discovery of which was reported in 1973 [55], are of
cosmological origin. The first cosmological model based on
explosions in active galaxy nuclei (AGN) was proposed in [84]. GRB
formation near a collapsing object due to neutrino– antineutrino
annihilation was investigated in [12]. Previously, the model of GRB
formation during collapses and explosions of supernovae was
considered in [23], where the following chain of reactions resulting
in gamma-photon production was considered:

\begin{equation}
 \label{eq1}
\tilde\nu+p \rightarrow n+e^+,
\end{equation}
\begin{equation}
 \label{eq2}
e^++e^- \rightarrow 2\gamma\,\,(0.5\,\, MeV),
\end{equation}
\begin{equation}
 \label{eq3}
n+p \rightarrow d+\gamma\,\, (2.3\,\,MeV),
\end{equation}
\begin{equation}
 \label{eq4}
d+p \rightarrow {\rm He}^3+\gamma\,\, (5.5\,\,MeV),
\end{equation}
\begin{equation}
 \label{eq5}
{\rm He}^3+{\rm He}^3 \rightarrow {\rm Be}^7+\gamma\,\,
(1.6\,\,MeV).
\end{equation}
 In brackets, we give
photon energies. The reactions with heavy nuclei were also
considered in [23]:

\begin{equation}
 \label{eq6}
\nu+(A,Z) \rightarrow  (A,Z+1)+e^-,
\end{equation}
\begin{equation}
 \label{eq7}
\tilde\nu+(A,Z) \rightarrow  (A,Z-1)+e^+.
\end{equation}

The starquake, the subsequent explosion, and the ejection of matter
from a nonequilibrium layer in the neutron-star crust discovered in
[22] are accom\-pa\-nied by the gamma rays induced by the fission of
superheavy nuclei. This scenario was proposed in [23] as an
alternative model of the galactic GRB origin. The total number of
various galactic models exceeds one hundred. Even now, after the
discovery of redshifts in optical afterglows, which became possible
due to the x -ray observations of GRB with the Beppo-SAX satellite
and their subsequent optical identification, the galactic models are
of more than simply historical interest because the origin of short
GRBs with the duration of less than 2 s remains uncertain. In [12],
it was obtained that the efficiency of transformation of
neutrino-flux energy $W_\nu \sim 6\cdot 10^{53}$ erg into the energy
of the $X$ - and $\gamma$ -ray burst amounts to a fraction $\alpha
\sim 6 \cdot 10^{–6}$ with a total GRB energy release $W_{X,\gamma}
\sim  3\cdot 10^{48}$ erg. The 3D numerical simulation of two
colliding neutron stars [86] and a hot torus around a black hole
[87] showed higher efficiency of $X$ - and $\gamma$ -ray production,
reaching 0.5\% in the first case and 1\% in the second case. In
part, the distinction with [12] can be associated with the geometry
more preferential for the neutrino-flux out- flow when the
annihilation rate increases in comparison with that estimated in the
spherical geometry [12]. Nevertheless, even in such an optimistic
variant, the GRB formation with a total energy yield in the $X$ -
and $\gamma$ -ray regions not exceeding $5 \cdot 10^{50}$ erg is
probable. This energy is insufficient for explaining the energy
resource of many GRBs because only the direct optical GRB radiation
can attain $10^{51}$ erg, and the isotropic flux in the gamma region
reaches $2.3\cdot 10^{54}$ erg for GRB 990123 with the redshift $z
\sim 1.6$ [2, 56]. To explain such a high observable energy release,
it is necessary to have a strong collimation, which, in turn, is
seriously restricted. Here, we discuss various GRB observational
data, analyze difficulties and problems of their interpretation in
the cosmological model, and consider physical restrictions on the
GRB models. In conclusion, we discuss the problems associated with
the interpretation of soft gamma repeaters (SGRs) as magnetars and
the possible relation between short GRBs and SGRs.

\section{Possible energy sources
of cosmic gamma-ray bursts}

For explaining the GRB radiation, the models of fireball, cannon
ball, and precessing jet are used. In these models, the key problem
associated with searching for the possibility of an enormous ($10^
{51} – 10^{54}$ erg) energy release over such a short (0.1–100 s)
time is avoided. From the various proposed models, we select the
following ones.

\smallskip

(i) The coalescence of two neutron stars or a neutron star and a
black hole of a stellar mass.

\smallskip

This mechanism was numerically investigated in [86, 87]. The
$\gamma$ rays are produced here due to the ($\nu,\tilde\nu$)
annihilation, and the energy yield proves to be insufficient for
explaining the most powerful GRBs even assuming a strong collimation
of the GRB radiation. The energy emitted only in the GRB 990123
optical afterglow [2, 56] exceeds the total radiation energy yield
in this model approximately by an order of magnitude.

\smallskip

(ii) Magnetorotational explosion.

\smallskip

The magnetorotational explosion proposed in [76] as the cosmological
GRB model was previously introduced in [15] for explaining supernova
explosions associated with a stellar core collapse. The 2D numerical
calculations of the explosion of a rotating magnetized gas cloud [5,
6] and the calculations of a magnetorotational supernova [71, 7, 8]
showed that the efficiency of transformation of the rotation energy
into explosive energy in both cases amounts to about 10\%. The
released energy is sufficient for explaining the explosion of a
collapsing supernova but proves to be insufficient for cosmological
GRB. The results of numerical simulation of a magnetorotational
explosion are shown in Fig. 1 [71].

\begin{figure}
\label{ardf1} \centerline
{\psfig{file=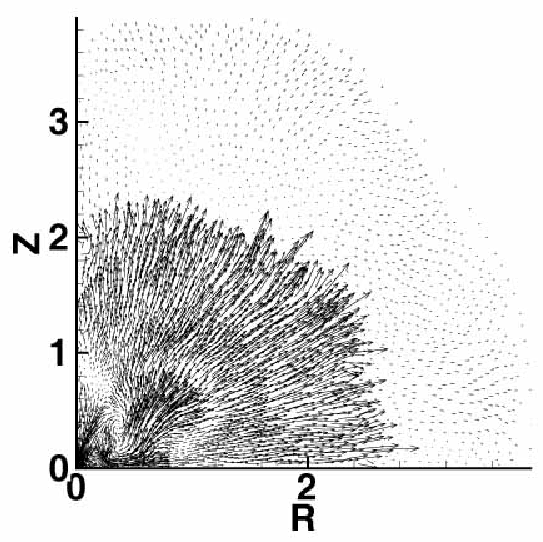,width=6.5cm,angle=-0}
 \psfig{file=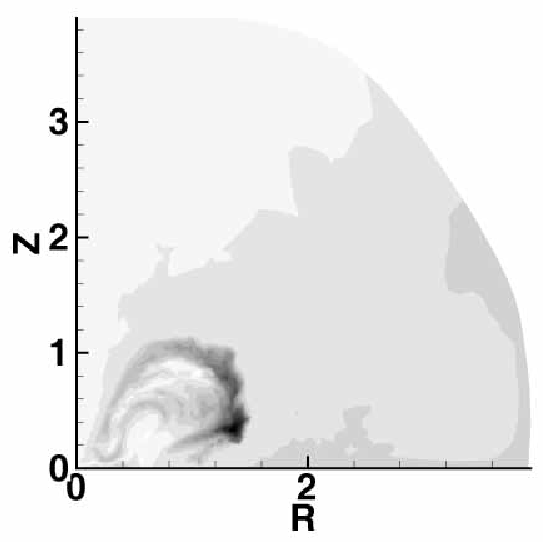,width=6.5cm,angle=-0}
 }
\caption{{{Velocity field (on the left) and the specific angular
momentum $v_\phi r$ (on the right) at the instant of time t = 0.191
s after switching on the magnetic field. High-angular-momentum
regions are indicated in the right-hand figure as the darker ones
[71].}}}
\end{figure}

\smallskip

(iii) Hypernova.

\smallskip

This model proposed in [76] assumes the possibility of explosion of
a very powerful supernova. Nowadays, this model is very popular
because it assumes that traces of supernova outbursts are found in
optical afterglows of several GRBs [98, 32, 100]. The radiation of a
strongly magnetized quickly rotating newly born neutron star for the
GRB origin was considered in [105]. Another model of hypernova
assumes the collapse of a massive nucleus, the formation of a black
hole with a mass $M_{bh} \sim 20 M_{\odot}$ surrounded by a massive
disk, the fast accretion of which results in the GRB phenomenon
[60]. This model seems to be the most promising now. To the
collapse-onset moment, massive stars collapsing with the formation
of black holes and GRBs were most likely [37, 104] in the Wolf–Rayet
state—a very bright massive compact star which lost its hydrogen
shell because of the outflow of matter during the prior evolution.
The SN Ic type of supernova, which is presumably observed at site of
GRBs and represents a product of the explosion of a massive star
devoid of a hydrogen shell, also points to this fact.

\smallskip

(iv) Magnetized disk around a (Kerr) rotating black hole (RBH).

\smallskip

This model is based on extracting the RBH rotation energy for the
GRB production due to its magnetic coupling with the RBH and the
surrounding accretion disk or torus [106].

\smallskip

(v) Dyadosphere.

\smallskip

In the model [88], the GRB originates from an explosive formation of
electron–positron pairs and the electromagnetic pulse from an
electrically charged black hole surrounded by a baryonic remnant.
This model is based on the vacuum explosion in the dyadosphere,
i.e., the region in which electric-field intensity supercritical
with respect to the formation of $e^{+}  e^{–}$ pairs is present.
The key problem here consists in the possibility of formation of
such a strongly charged BH (see also [74]).

\smallskip

(vi) A shock wave beyond the neutron star formed after the supernova
explosion in a double system was considered as the GRB source in
[51].

\smallskip

(vii) An exotic model of production of GRBs from superconducting
strings was proposed in [13].

\smallskip

(viii) The phase transition with the formation of a quark (strange)
star.

\smallskip

The energy release at the transition of a neutron star into a stable
equilibrium state of a quark (strange) star was considered in [14]
as the solution to the problem of cosmological GRB. This model could
explain the relation of GRB to supernova explosions with the
formation of a neutron star and the subsequent enormous energy
release during its transition into the state of a quark star leading
to the GRB. An attractive property of this model is the possibility
of obtaining an arbitrary delay time between the SN explosion and
the GRB, which is associated with the transition of hadrons in a
quark state and very strongly depends on various parameters. In this
model, the delay time can be arbitrarily large so that the majority
of SN explosions should result in no GRBs according to observations
if this time exceeds the Hubble lifetime of the Universe. Free
quarks should not virtually interact, and the density in the case
that noninteracting quarks enter an energetically favorable state is
very uncertain [14]. In Figs. 2 and 3, the models of neutron and
quark (strange) stars are shown on the mass–radius plane, and a
broad spread in theoretical predictions is seen. A greater spread in
the properties of quark stars in Fig. 3 is obtained in [4] within
the framework of the same model of quark matter as in Fig. 2, but
for a wider variation of parameters.

In the presence of considerable theoretical problems, only
observations can settle the question about the possible existence of
quark stars. Even in the presence of theoretical uncertainties, it
is impossible to obtain a neutron (purely hadronic) star with a
radius essentially smaller than 10 km. The observational discovery
of compact stars with such small radii would be evidence against
their purely hadronic composition.

\begin{figure}
\centerline{ \psfig{file=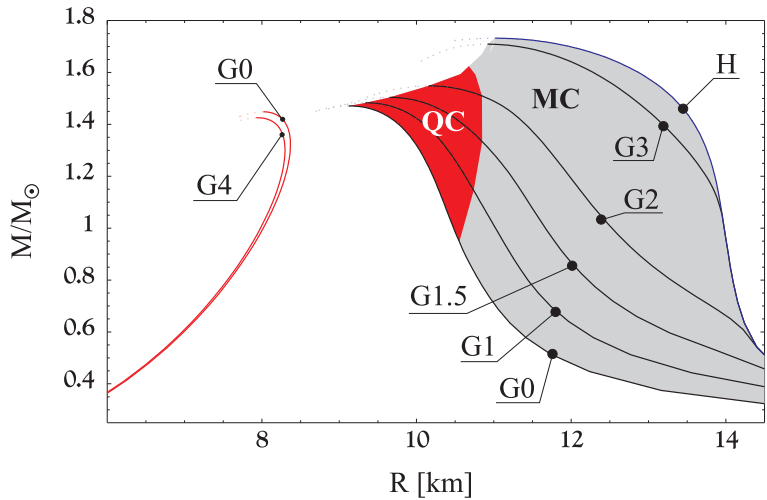,width=10cm,angle=-0} }
\caption{Mass–radius relation for pure “strange” quark stars (on the
left) and hybrid stars (on the right). The G0–G4 models for hybrid
stars correspond to different parameters of the model. H is the pure
hadronic star, QC is the quarkcore star, and MC is the mixed-core
star [101].} \label{quark1}
\end{figure}

\begin{figure}
\centerline{\psfig{file=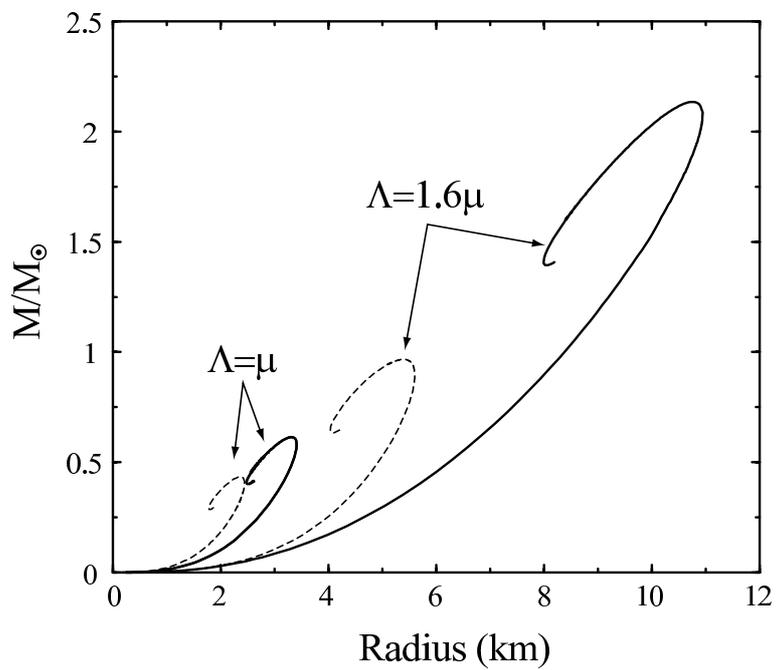,width=10cm}}
\caption{Mass–radius dependence for quark stars with $\Lambda/\mu$ =
1.6 and 1. Dashed lines represent the models with a weak coupling,
which takes place for the same choice of renormalization scales [4].
} \label{quark2}
\end{figure}

\section{Statistical properties of cosmic gamma-ray bursts}

The conclusion about the cosmological origin of GRBs is based on the
analysis of their statistical properties and optical-afterglow
spectra with strongly redshifted lines.

Statistical arguments for the cosmological origin of GRBs are
associated with an observed isotropy of the GRB distribution over
the sky together with a strong deviation of the $\log N\,–\,\log S$
distribution (or an equivalent) from the Euclidean homogeneous
distribution with a slope of 3/2. The homogeneous GRB distribution
over the celestial sphere was obtained first in the experiment KONUS
[62] and then confirmed in the experiment BATSE [68] in which more
than 3000 GRBs were detected in total during nine years (1991–2000)
of observation in space. The results obtained in the experiment
KONUS are shown in Fig. 4; in this experiment, 143 GRBs [63] were
detected in 384 days.

\begin{figure}
\centerline {\psfig{file=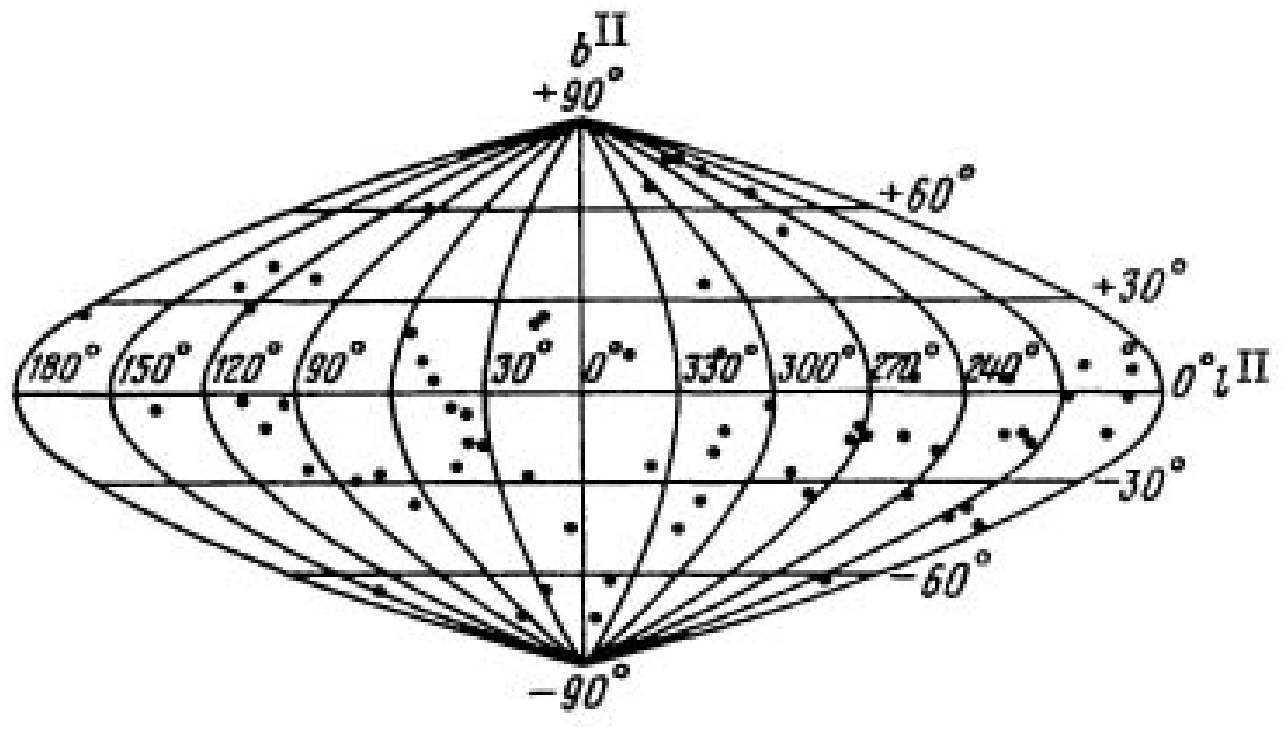,width=6.5cm,angle=-0}
 \psfig{file=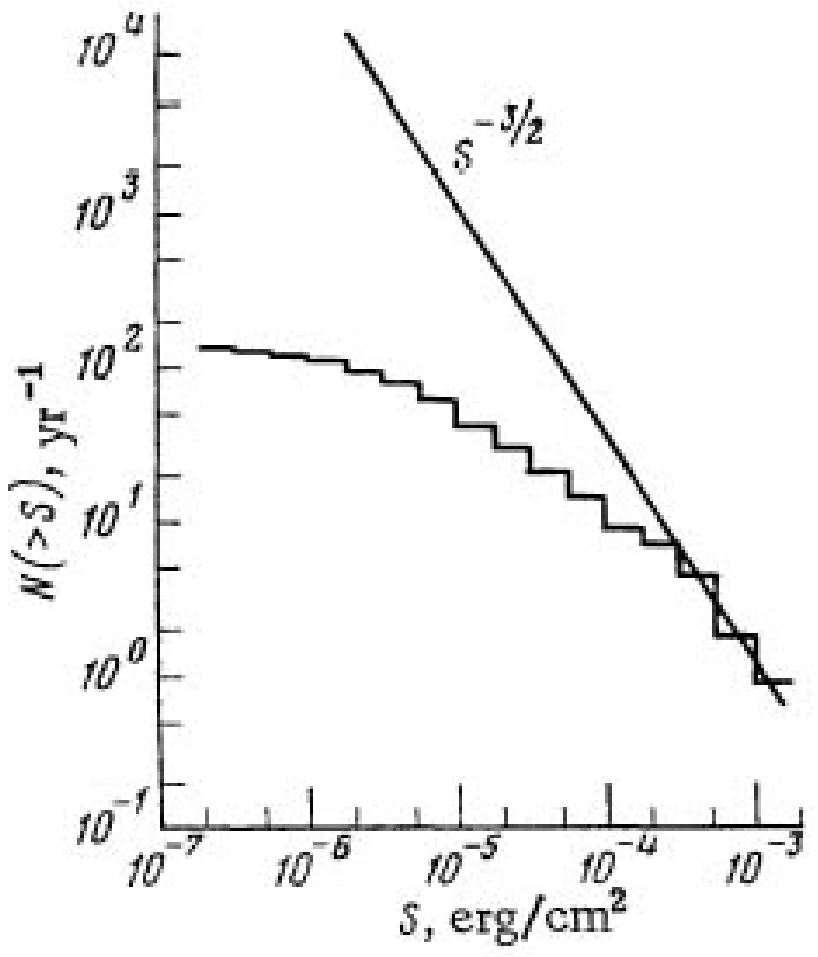,width=6.5cm,angle=-0}
 }
\caption{GRBs detected in the experiment KONUS. On the left: the GRB
position at the celestial sphere represented in the galactic
coordinates $l^{II}$ and $b^{II}$. On the right: $\log N\, -\, \log
S$ distribution [62].} \label{grbfig1}
\end{figure}

The authors of [62] assumed that the properties observed for the
 $\log N\,–\,\log S$
 curve are associated with various selection effects and
that the true distribution of GRB sources in space is homogeneous.
The selection effects in the experiment KONUS taken into account in
[42] led to the average value $<V/Vmax> = 0.45 \pm 0.03$; here, 0.5
corresponds to a homogeneous distribution of sources. The KONUS data
were obtained in 1978–1980 under conditions of a constant background
during the flight to Venus. The similar analysis of the BATSE data
[91] obtained at a circumterrestrial orbit under conditions of a
variable background gave $<V/V_{max}>=0.334 \pm 0.008$.
Figures 5 and 6 display the statistical characteristics
of the BATSE data [68, 35] and the number of
GRBs here is more than an order of magnitude higher.
These two results seem to contradict each other. In the
KONUS experiment, the sensitivity is approximately
3 times lower than in the BATSE experiment, and the
deviations from a homogeneous distribution with
$<V/V_{max}>=0.5$
 began much earlier [35]. We show the
BATSE–PVO joint results in Fig. 7 from [35]. The PVO
data on the brightest GRBs show good spatial uniformity
for GRBs with a slope of 3/2.
In [92], M. Schmidt made a detailed statistical analysis
and processed the BATSE data by dividing the
GRBs into four classes according to the degree of spectrum
hardness; he calculated $<V/V_{max}>$ for each class
individually. The results of this investigation are listed
in Table 1, where $\alpha_{23}$ determines the spectrum slope
obtained from the counts in the second and third
BATSE channels with the energy ranges of 100–300 keV
and 50–100 keV, respectively. In the column
“obs,” we list the observables corrected taking into
account statistical errors in count peaks and, in the column
“corr,” we give the same data assuming the existence
of a luminosity–degree-of-hardness correlation.

\begin{table}
\caption{1. Dependence of ${<V/V_{max}>}$ on the degree of
hardness for 1391 GRB [92]}
%\begin{center}
\medskip
\begin{tabular}{ccccc}
\hline\noalign{\smallskip} {number}& {$<\alpha_{23}>_{obs}$}&
{${<V/V_{max}>}_{obs}$}& {$<\alpha_{23}>_{corr}$}&
{${<V/V_{max}>}_{corr}$}\\
\noalign{\smallskip}
\hline\\
 348 & $-2.55$ & $0.468 \pm 0.017$ & $-2.33$ & $0.421$ \\
 348 & $-1.84$ & $0.309 \pm 0.016$ & $-1.79$ & $0.325$ \\
 347 & $-1.47$ & $0.299 \pm 0.016$ & $-1.47$ & $0.344$ \\
 348 & $-1.04$ & $0.270 \pm 0.015$ & $-1.10$ & $0.256$ \\
\hline\\
\end{tabular}
%\end{center}
\end{table}
In the cosmological model, it was possible to expect
a lower value of $<V/V_{max}>$
 for softer GRBs in the case of
a homogeneous set because, with an increase in the redshift,
the spectrum becomes softer. The result proved to
be the opposite—soft sources had a greater value of $<V/V_{max}>$
than the hard ones: 0.47 and 0.27, respectively.

\begin{figure}
\centerline {\psfig{file=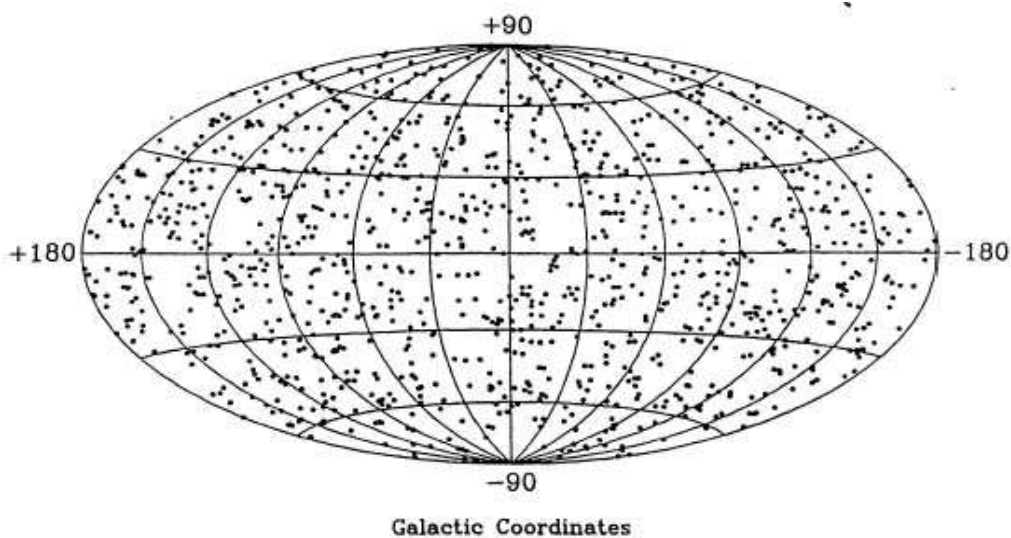,width=14cm,angle=-0}
 }
\caption{Distribution of 1121 GRBs over the celestial sphere in
galactic coordinates detected by BATSE over three years. No
clusterization
or anisotropy is observed [35].} \label{grbfig2}
\end{figure}

\noindent
In [92], it was assumed that the increase in the GRB
power with the degree of spectrum hardness is so significant
that it outweighs the opposite effect caused by
the redshift in a homogeneous sample of objects.
Another explanation can be that the sample of softer
GRBs is more complete because of selection effects.
Figures 8 and 9 illustrate a possible key role of the
selection effects (incompleteness of the data set and
statistical errors in estimates of luminosity in the presence
of threshold). The incompleteness of data affects the
distribution of such well-investigated objects as G
stars of the Sun type. It is possible to expect still greater
effects when detecting such short transients as GRBs.
The comparison of average values of $V/V_{max}$ as functions
of the detection threshold for G stars and GRBs in
Fig. 8 from [41] shows their qualitative agreement.

\begin{figure}
\centerline {\psfig{file=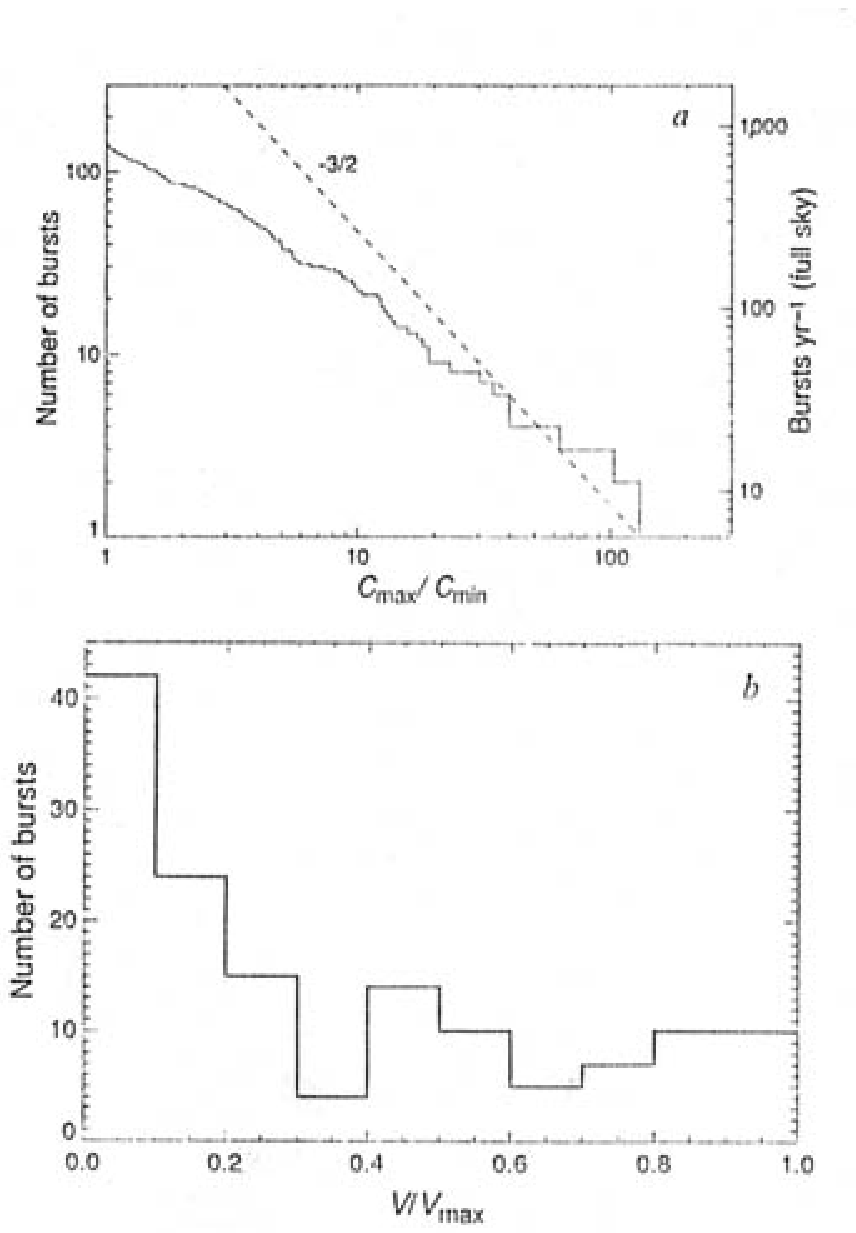,width=11.5cm,angle=-0}
 }
\caption{(a) Integrated distribution of the GRB number as a
function of the maximal counting rate for 140 GRBs. For a
homogeneous distribution of sources, the power law with an
exponent –3/2 is expected. The total detection rate for bursts
amounts to $\sim 800$ events per year. (b) The distribution of
V/Vmax is given for 140 GRBs. An average $V/V_{max}$ is equal
to 0.348$\pm$0.024 [68].} \label{grbfig3}
\end{figure}
\noindent The effect of statistical errors in the presence of a
detection threshold is investigated in [17]. The
$V/V_{max}$ distributions with average statistical errors at the
level of ten thresholds are shown in Fig. 9a for the normal
distribution of counts and in Fig. 9b for the similar
distribution over their logarithms. The distribution in
Fig. 9a is similar to the BATSE distribution in Fig. 6.

\begin{figure}
\centerline {\psfig{file=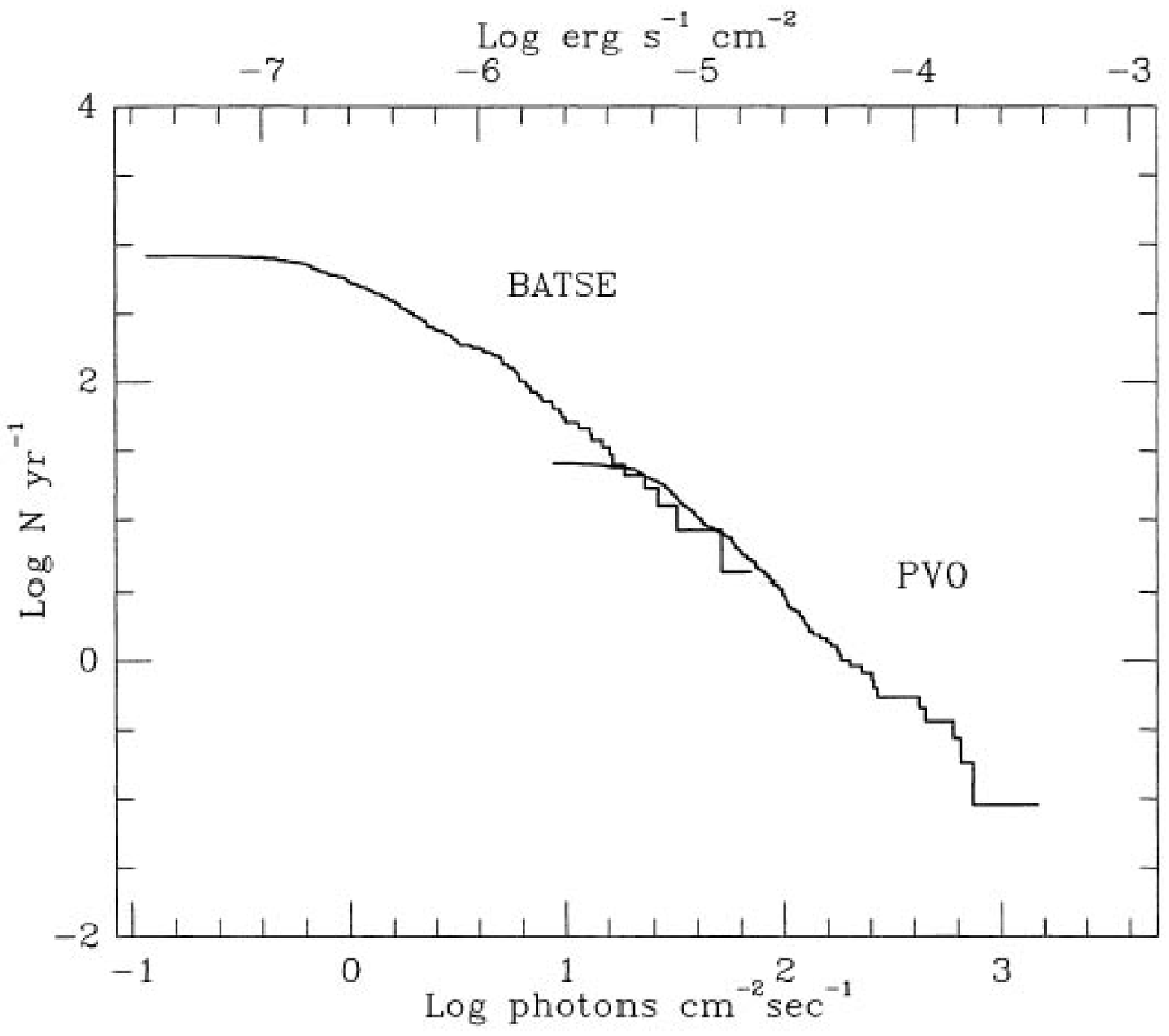,width=14cm,angle=-0}
 }
\caption{Distribution $\log N\, -\, \log P$
 for a combined set of
BATSE–PVO data. The distributions match well in the dataoverlap
region. It can be seen that the PVO data containing
a larger number of powerful bursts than the BATSE data and
obtained for longer time well follow the power law with an
exponent –3/2 for powerful bursts [35].} \label{grbfig4}
\end{figure}

\begin{figure}
\centerline {\psfig{file=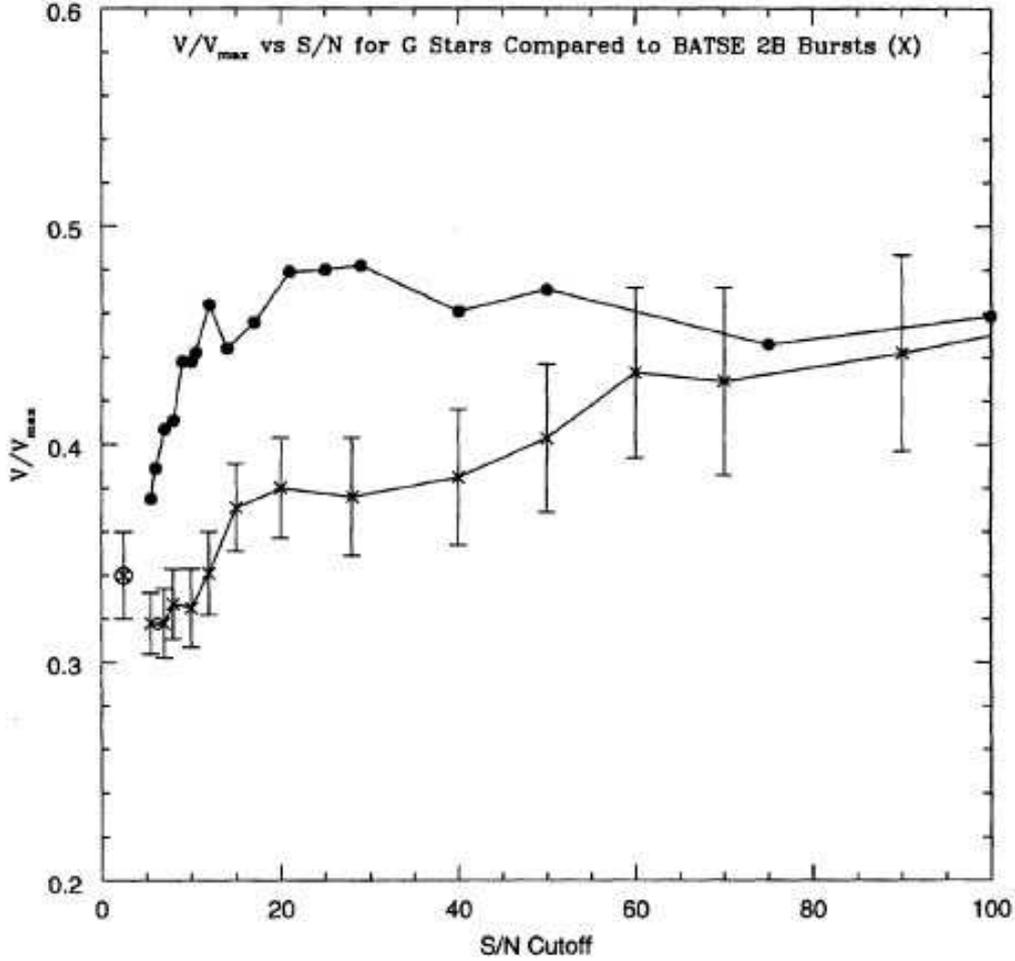,width=14cm,angle=-0}
 }
\caption[h]{Average value of $V/V_{max}$ as a function of the minimal
brightness of the objects taken into account for G stars
(closed circles) and GRBs from BATSE 2B catalogue
(crosses with the indication of the 1$\sigma$-error interval). For G
stars, the error intervals are less than the circle sizes. The
crosses in circles give a value of average
$V/V_{max}$ from the
BATSE 1B catalogue. The limit of the GRB-detection
established with BATSE corresponding to 5.5$\sigma$. For G stars,
the limiting value $m_v=10$ corresponding to S/N = 5.5 was
used in this analyses. Note that $<V/V_{max}>$. is close to the uniform
value 0.50 both for bright G stars and for bright GRBs.
By taking weaker sources into account, $<V/V_{max}>$ deviates in
the lower-value side. For G stars, this fact is associated with
an incompleteness of the catalogue of weak sources [41].}
\label{grbfig5}
\end{figure}

\begin{figure}
\centerline {\psfig{file=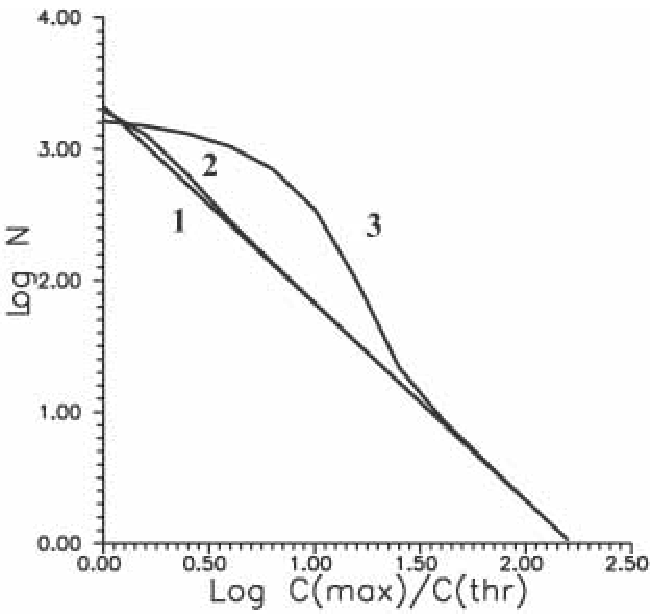,width=7cm,angle=-0}
 \psfig{file=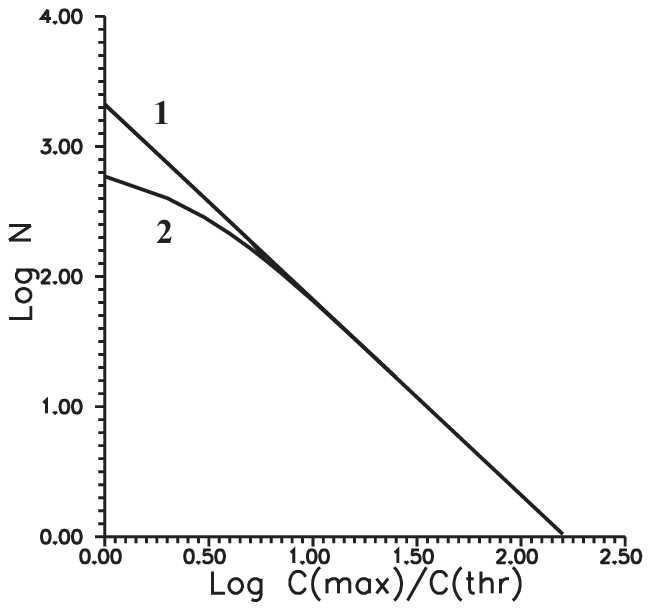,width=7cm,angle=-0}
 }
 \caption{(a) Curve $[\log N - \log C(max)/C(thr)]$
  in the presence
 of statistical errors normally distributed with an average error $\Delta_1$ in
detection-threshold units; 1, is the straight line with a slope
of –3/2 corresponding to $\Delta_1=0$; 2, is the curve with $\Delta_1=1$; and 3, is
the curve with $\Delta_1=10$. (b) The same as in Fig. 9a for the normal
logarithmic distribution of errors; $\Delta$ determines the average error
as a spread of the threshold-number logarithm; 1, is the straight
line with a slope of –3/2 corresponding to $\Delta = 0$; 2, is the
curve with $\Delta = 1$; C(max) is the peak burst intensity; and C(thr)
is the value of detection threshold [17].}
   \label{stat}
\end{figure}

\section{Optical afterglows and redshifts}

The x-ray afterglow discovered by the Italian Beppo-SAX satellite
made it possible to identify GRBs optically and to obtain their
spectra. These spectra showed the presence of a large redshift z,
reaching 4.5, which points to the cosmological origin of GRBs and an
enormous energy release. In most cases, the redshifts were measured
for very weak parent galaxies. The list of certain redshifts is
given in Table 2, where the redshift data taken from [34] are
supplemented with the total radiated GRB energy [19]. In the table,
we listed the numbers of triggers and total energy fluxes of GRBs
from the 4B catalogue [75] and total energy fluxes from other
sources. The location of optical afterglows in parent galaxies is
given in [26]. The enormous energy yield in a short time (from 0.1 s
to several hundreds of seconds) is the primary difficulty in the
cosmological interpretation. In certain cases, for example, in GRB
990123, the burst isotropic energy ($2.3\cdot 10^{54}$ erg) exceeds
the rest energy of the Sun. All of the mechanisms discussed above,
with the exception of the Dyadosphere model of Ruffini and
collaborators, could give an energy release less than one percent of
this quantity. In the case of Dyadosophere the Blackholic energy
could fulfill the above requirement

\begin{table}[hp]
\caption{Redshifts of parent galaxies (Z) and GRB total energy
fluencies (F), June 2001} \small{
\begin{center}
{\begin{tabular}{lllllll} \hline\noalign{\smallskip}

 Trigger&
  GRB &
  $R$ mag &
  Z &
  Type $^a$&
  F$^e$&
   Ref.\\
number&&&& & erg/cm$^2$&\\
\noalign{\smallskip} \hline \noalign{\smallskip}

& 970228     &   25.2 &  0.695   & e & 10$^{-5}$&\cite{hc97}  \\
6225& 970508     &   25.7 &  0.835   & a,e & $3.5\cdot 10^{-6}$(3+4)&  \\
6350& 970828     &   24.5 &  0.9579  & e & $7 \cdot 10^{-5}$&\cite{gg98} \\
6533& 971214     &   25.6 &  3.418   & e & $ 10^{-5}$(3+4)&   \\
6659& 980326     &   29.2 &$\sim$1?  &   & $6.3 \cdot 10^{-7}$(3+4)& \\
6665& 980329     &   27.7 &$<$3.9    & (b)&$7.1\cdot 10^{-5}$(3+4)& \\
6707& 980425 $^c$&   14   &  0.0085  & a,e&$4.4 \cdot 10^{-6}$&\cite{gv98}  \\
6764& 980519     &   26.2 &          &  &$9.4 \cdot 10^{-6}$(all 4) &    \\
& 980613     &   24.0 &  1.097   & e &$1.7\cdot 10^{-6}$&\cite{gn112} \\
6891& 980703     &   22.6 &  0.966   & a,e&$5.4\cdot 10^{-5}$(3+4)&\cite{gn126} \\
7281& 981226     &   24.8 &          & &$2.3\cdot 10^{-6}$(3+4)&    \\
7343& 990123     &   23.9 &  1.600   & a,e &$5.1 \cdot 10^{-4}$&\cite{gn224}\\
7457& 990308 $^d$&$>$28.5 &          &    &$1.9\cdot 10^{-5}(3+4)$ \\
7549& 990506     &   24.8 &  1.30    & e  &$2.2 \cdot 10^{-4}$&\cite{gn306}  \\
7560& 990510     &   28.5 &  1.619   & a &$2.6 \cdot 10^{-5}$&\cite{gn322} \\
& 990705     &   22.8 &  0.86    & x&$\sim 3 \cdot 10^{-5}$ &\cite{mas00}    \\
& 990712     &   21.8 &  0.4331  & a,e&&  \\
& 991208     &   24.4 &  0.7055  & e&$\sim 10^{-4}$ &\cite{gn450} \\
7906& 991216     &   24.85&  1.02    & a,x&$2.1\cdot 10^{-4}$(3+4)& \\
7975& 000131     &$>$25.7 &  4.50    & b &$\sim 10^{-5}$& \cite{gn529}  \\
& 000214     &        &0.37--0.47& x&$\sim 2 \cdot 10^{-5}$&\cite{gn557}    \\
& 000301C    &   28.0 &  2.0335  & a  & $\sim 4\cdot 10^{-6}$ &\cite{gn568}\\
& 000418     &   23.9 &  1.1185  & e&$1.3\cdot 10^{-5}$&\cite{gn642}  \\
& 000630     &   26.7 &          &   &$2\cdot 10^{-6}$&\cite{gn736}  \\
& 000911     &   25.0 &  1.0585  & e &$5 \cdot 10^{-6}$&\cite{gn791}  \\
& 000926     &   23.9 &  2.0369  & a &$2.2 \cdot 10^{-5}$&\cite{gn802}  \\
& 010222     &$>$24   &  1.477   & a& The brightest on  &\cite{gn959} \\
&&&&&BeppoSAX&\\
\hline
\end{tabular}

\noindent$^a$ e is the emission line, a is the absorption line, b is the continuum break, and x are the x-rays.

\noindent$^c$ Relation of this galaxy with SN or GRB is not quite reliable.

\noindent$^d$ Relation of optical transient with this GRB is unreliable.

\noindent$^e$ Number of BATSE channel with a maximal flux is indicated in parentheses [75];
otherwise, the estimates of the total energy flux are
given from other references.\\
}
\end{center}
} \label{Tab1}
\end{table}

\subsection{Collimation}

To avoid an enormous production of energy, it is assumed that the
GRB radiation is collimated. In the cannon-ball model [32], it is
assumed that a whole radiating object moves relativistically with
a large factor $\Gamma \approx 10^2\,-\,10^3$ that leads to a
collimation solid angle $\Omega \approx 10^{-4}\,-\,10^{-6}$.
 The analysis of the GRB collimation was made in [85]. A
strong collimation-angle restriction follows from the analysis of
the probability of the appearance of orphan optical afterglows,
which are most likely poorly collimated or not collimated at all.
The absence of certain rapidly alternating orphan optical objects
results in the following restrictions. In the case of isotropic
GRBs, it was expected to detect $\sim 0.2$ orphan afterglows so
that the absence of such objects assumes that
$\Omega_{opt}/\Omega_{\gamma} \ll 100$, which suffices for ruling
out an especially strong collimation. The similar investigation
using the data on afterglows in the radio-frequency band and the
variable radio sources made in [79] leads to the restriction on
the collimation angle $\theta_{\gamma} \geq 5^{\circ}$. Because
the afterglows in the radio-frequency band occur at the
nonrelativistic stage of evolution of a GRB remnant, it is
expected to be isotropic, and the restrictions following from the
absence of radio orphans at $\Omega_r/\Omega_{\gamma}$ establish a
limit on the radiation collimation $\Omega_{\gamma}$ of the GRB
itself.

The comparison of redshifts and total energy fluxes from Table 2
shows no correlation between the distance and observable flux (see
Fig. 10). This fact is usually explained by a strong collimation
leading to a wide spread in observed fluxes due to observations of
a narrow beam under different angles of view. If the collimation
were associated with the relativistic motion of a source [32], the
strong correlation between the observed radiated energy and the
GRB duration would be detected in this case: powerful GRBs should
have been shorter. The absence of such a correlation allows one to
rule out the models based on the ultrarelativistic motion of
sources.

\begin{figure}
\centerline {\psfig{file=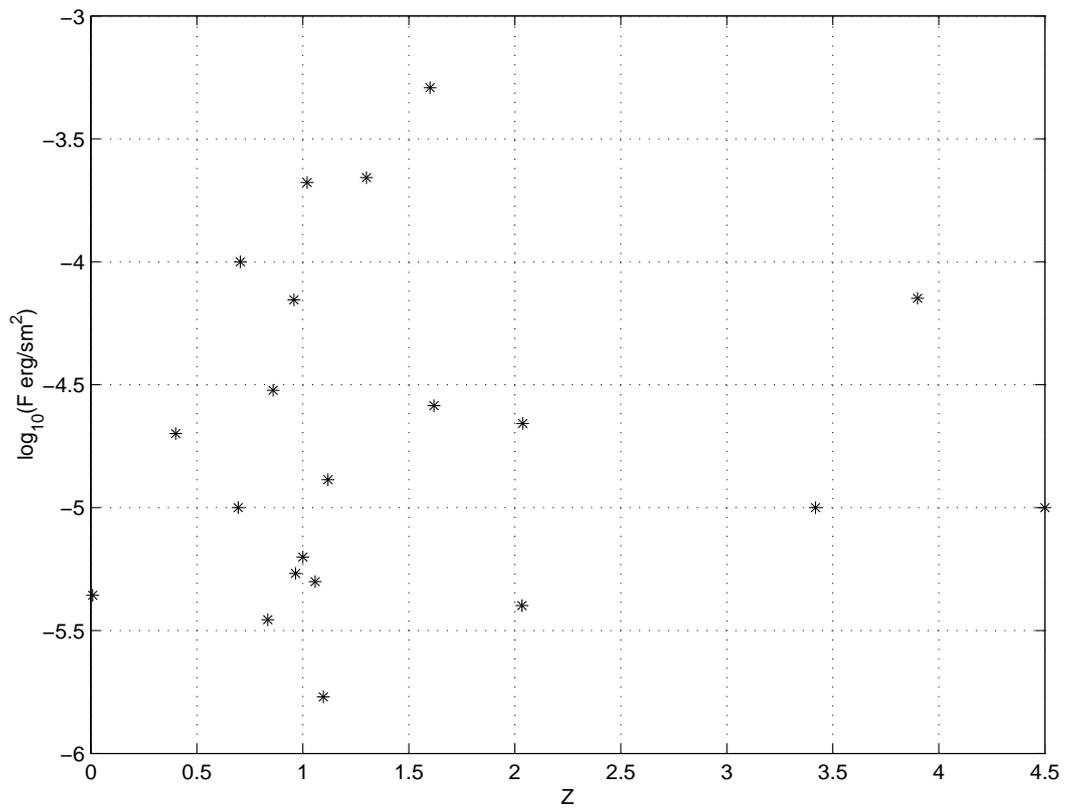,width=14cm,angle=-0}
 }
\caption{Total observed energy fluence F as a function of the
redshift z for GRBs from Table 2 [20].} \label{bkz}
\end{figure}

\subsection{Prompt Optical Afterglows}

The optical afterglow of GRB 990123 was successfully observed 22 s
after the beginning of its detection in the gamma region [3, 2].
GRB 990123 was detected by BATSE in 1999 on January 23.407594. The
burst was strong with duration $\ge$100 s, involving several peaks
(see Fig. 11), and revealed a strong spectrum evolution. The T50
and T90 durations at the 50 and 90\% energy-flux levels were equal
to 29.82 ($\pm$0.10) and 63.30 ($\pm$0.26) s, respectively. The
peak of optical brightness at a level of 8.95$^m$ was achieved
within 30 s after the burst onset, and the brightness decreased
down to 14.5$^m$ within 95 s. Thus, the optical-radiation peak
almost coincided, with a certain delay, with the gamma ray peak.
The observed optical luminosity, which was related to the redshift
$z = 1.61$, achieved $L_{opt} \approx 4 \cdot 10^{49}$
 erg/s, which is brighter by almost five orders of magnitude
than the optical luminosity of any supernova. The energy release
in the direct optical radiation reaches $10^{51}$ erg, and an
isotropic gamma-ray flux exceeds $2.3 \cdot 10^{54}$ erg, which is
more than the Sun rest energy [2, 56]. Another bright afterglow
was observed for GRB 021004 ($15^m$, $z = 2.3$), GRB 030329
($12.4^m$, $z = 0.168$), and GRB 030418 ($16.9^m$). In brackets,
we list the brightest visual magnitudes and redshifts.

The GRB 030329 afterglow observed at many observatories is most
interesting (see, for example, [89]). Features in the afterglow
spectrum showed the presence of a supernova [100]. Figure 12 shows
the .-ray observations for GRB 030329 in the KONUS– WIND
experiment, and Fig. 13 taken from [82] shows the light curve for
the optical afterglow obtained at the Crimean observatory.

\begin{figure}
\centerline {\psfig{file=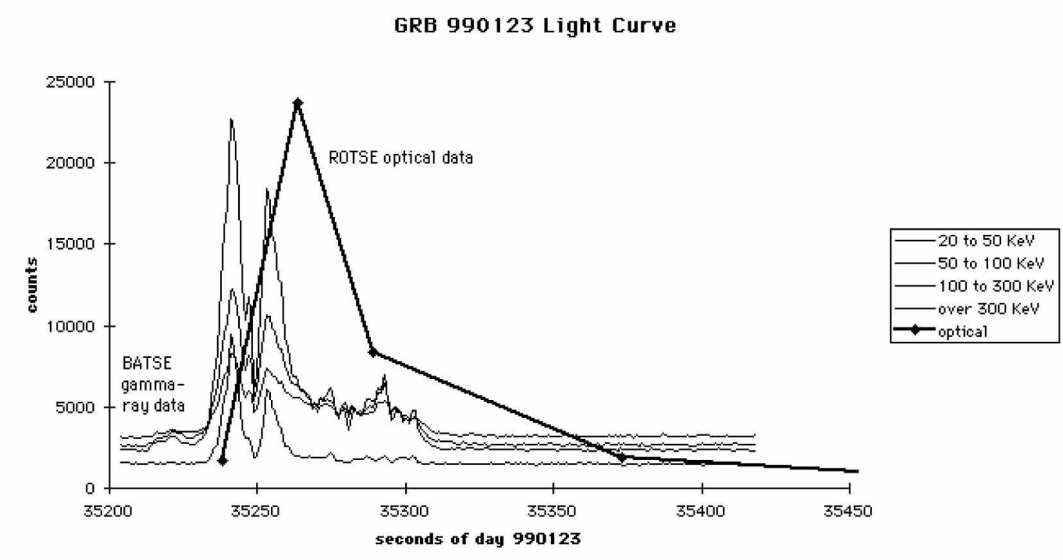,width=14cm,angle=-0}
 }
\caption{Thin lines reproduce the gamma-ray profile from the BATSE
data with the resolution of 1024 ms in various energy channels.
The thick line connects some first dots of ROTSE optical
observations, which began within 22 s after the detection. This
burst is the brightest one in the optical range [82].}
\label{opt99}
\end{figure}

\begin{figure}
\centerline {\psfig{file=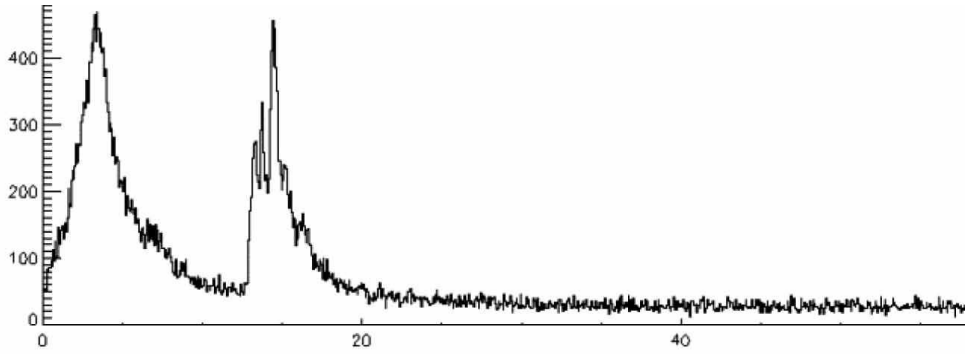,width=13cm,angle=-0}
 }
\caption{GRB 030329 in gamma rays: KONUS–WIND (UT)11:37:29. The
total energy fluence is equal to $1.2 \times 10^{-4}$~erg/cm$^2$,
the duration is 50 s, and the maximal flux is $2.5 \times 10^{-5}$
erg/(cm$^2$s); one of the brightest bursts in the gamma region
[82].} \label{gam03}
\end{figure}

\begin{figure}
\centerline {\psfig{file=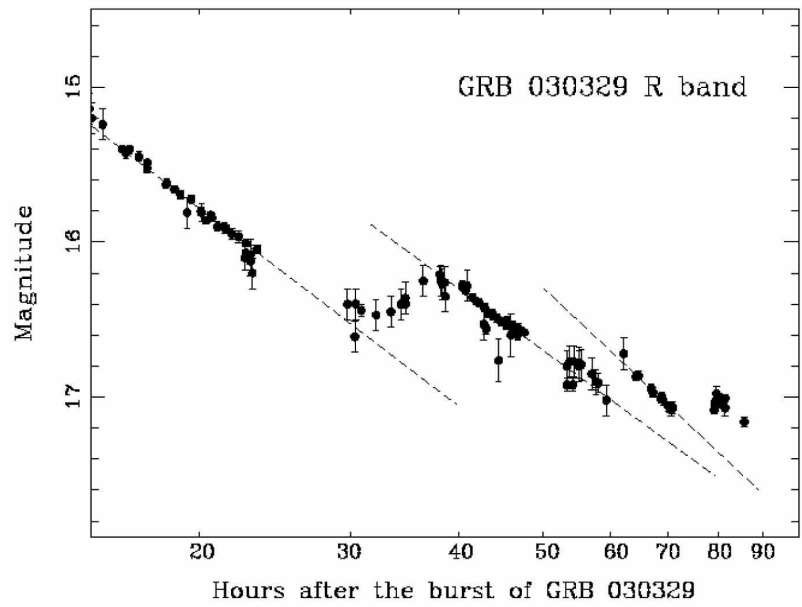,width=12.5cm,angle=-0}
 }
\caption{Hour-scale variation of the GRB 030329 optical afterglow
[82].} \label{opt03}
\end{figure}

\begin{figure}
\centerline {\psfig{file=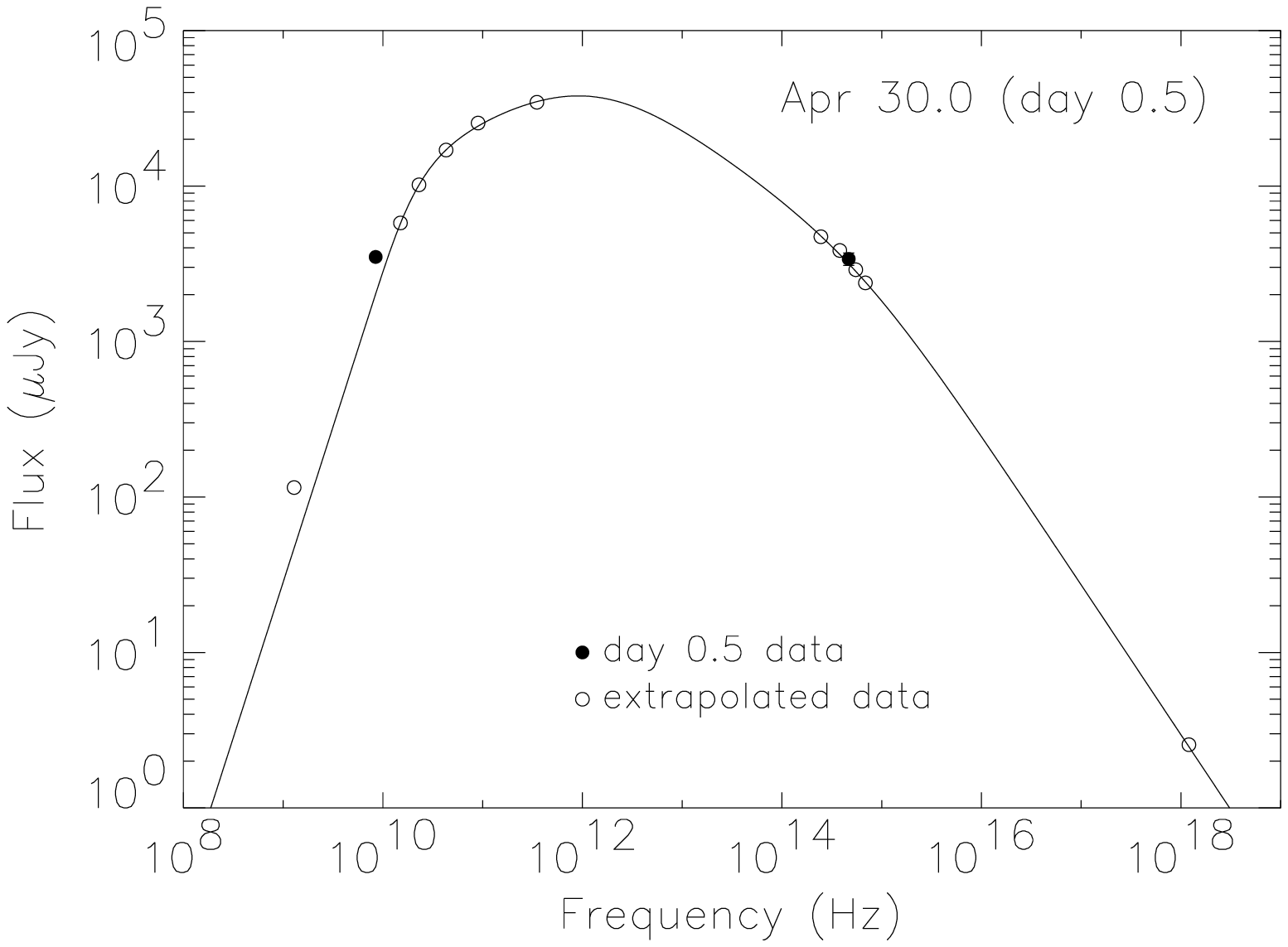,width=7cm,angle=-0}
\psfig{file=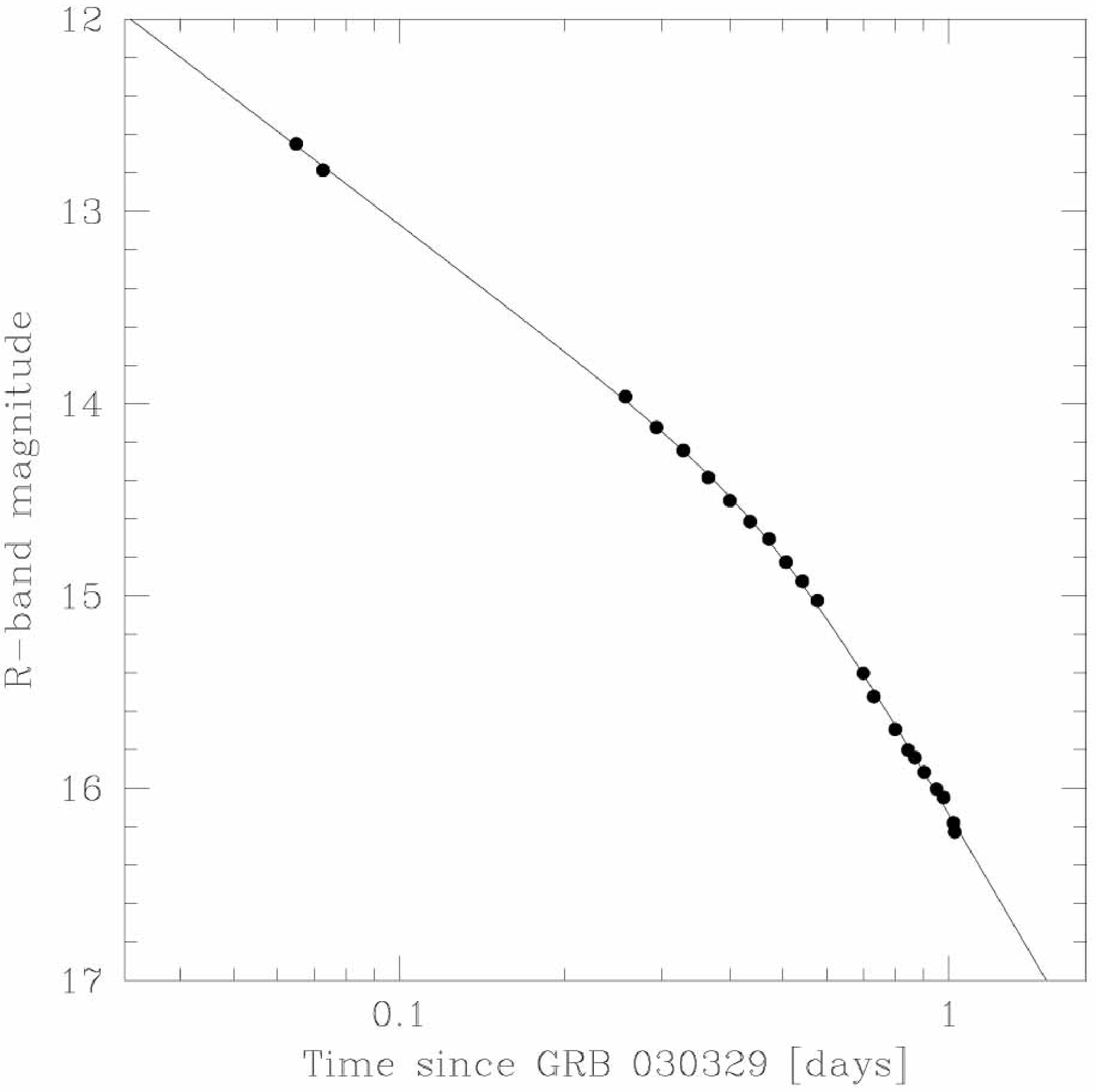,width=7cm,angle=-0}
 }
\caption{On the left: the GRB 030329 afterglow spectrum. This
spectrum covering a wide region is obtained within 0.5~days after
the GRB. Closed circles correspond to the measurements obtained
nearby the nominal time, and open circles correspond to the
extrapolation to the nominal time assuming the evolution in a
constant-density medium. The simple fit of this broad spectrum
gives the following parameters: the synchrotron self-absorption
frequency $\nu_a \sim 25\,$GHz, the spectral-peak frequency
$\nu_m\sim 1270\,$GHz, the cooling frequency $\nu_c\sim 6.2\times
10^{14}\,$Hz, the spectral-peak flux $f_m\sim 65\,$mJy, and the
electron-energy index $p \sim 2$. The following physical
parameters correspond to these values: the explosion energy $E\sim
5.7\times 10^{51}\,$erg, the surrounding-gas density $n \sim 5.5$
atom cm$^{–3}$, the electron-energy fraction $\epsilon_e\sim
0.16$, and the magnetic-field-energy fraction $\epsilon_B\sim
0.012$. On the right: the light curve of the GRB~030329 optical
afterglow in the $R$ region of the spectrum beginning
approximately after one day from the GRB moment. The observation
errors are less than the data point sizes [83].} \label{optic03}
\end{figure}

\begin{figure}
\centerline {\psfig{file=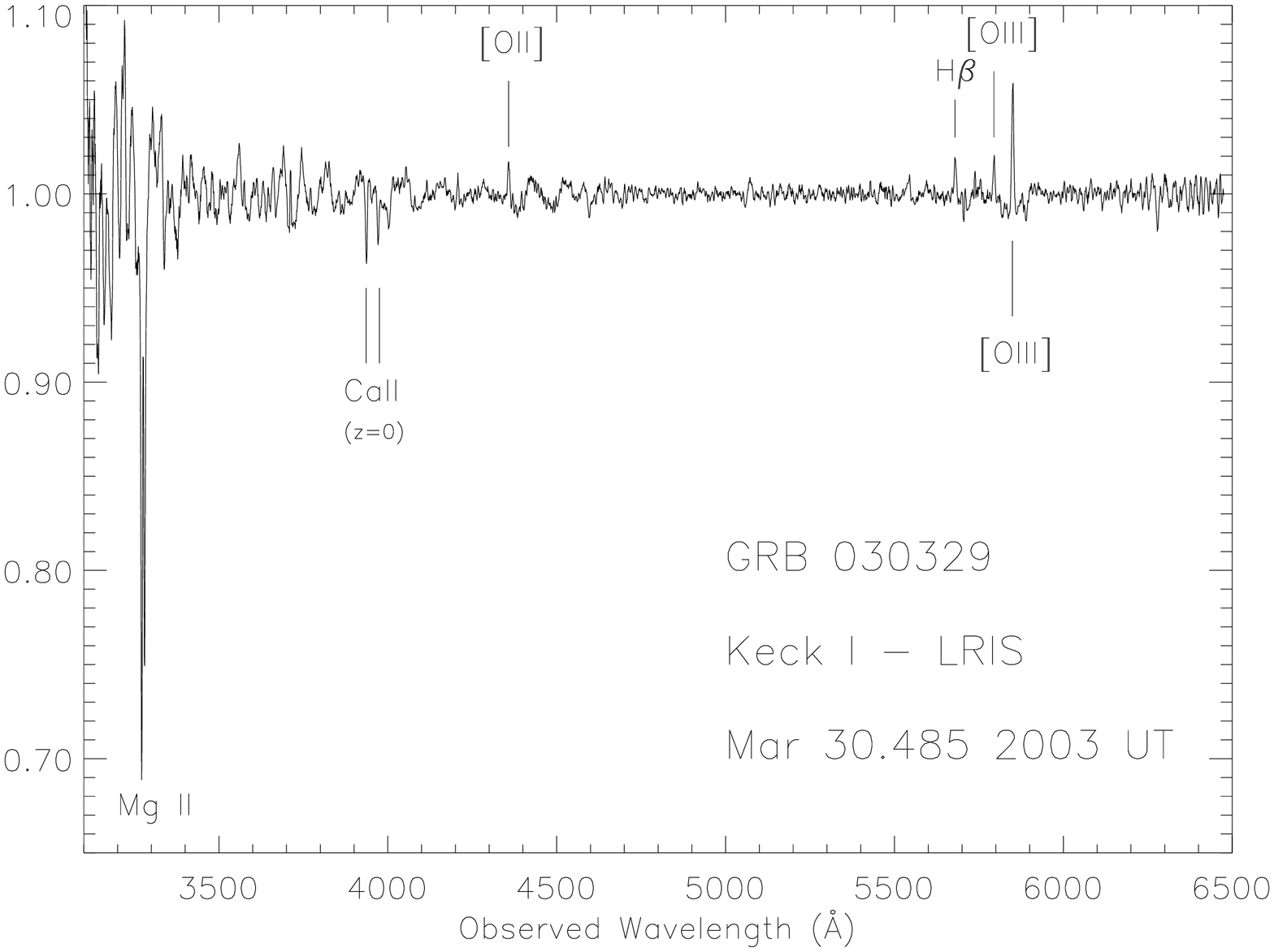,width=14cm,angle=90}
 }
\caption{Spectrum of optical afterglow obtained with the telescope
Keck I using the spectrometer with an effective resolution of 4.2
\AA. The exposition in the afterglow observation amounted to 600~
s. The identified narrow emission lines [O~II], H$\beta$, [O~III],
and the lines of Mg~II absorption correspond to an average
redshift $z = 0.169 \pm 0.001$, which makes GRB~030329 to be a
cosmological GRB with the smallest redshift. The identified
emission lines are typical for star-formation regions in galaxies,
while the absorption lines are caused by gas in the galaxy disk.
The identified line Ca~II at $z \approx 0$ is hypothetically
associated with the absorption in a cloud inside our Galaxy [83].
  }
\label{spec03}
\end{figure}

\section{GRB and supernovae}

The observational indications of the relation to supernovae are
found for the following GRBs: GRB 980425 ($z = 0.0085$, 40 Mpc),
GRB 980326 ($z = l$), GRB 011121 ($z = 0.365$), GRB 020405 ($z =
0.695$), and GRB 030329 ($z = 0.169$). The light curves for
certain optical afterglows have “a red bump” for 15–75 days, which
may be induced by an accompanying supernova explosion (see [98],
Fig. 16 from [27], and Figs. 18 and 19 from [43]). The indications
of the relation between GRBs and supernova explosions were
obtained in the observation of GRB 980425, which coincided in the
sky with SN 1998bw. This supernova proved to be abnormally bright
in comparison with other SNe (such as SN Ib/c) and had peculiar
spectra and unusually high radio-luminosity at initial stages
[78]. It exploded in a nearby galaxy with a much smaller redshift
than those observed in the optical afterglows of other GRBs from
Table 2, which points to a low energy release in GRB 980425. It is
impossible to rule out that the SN and GRB approximate coincidence
in time and projection to the celestial sphere is accidental, and
the distances to them are very different. Nevertheless, even given
the obvious peculiarity of nearby GRB 980425 with its relation to
SN 1998bw, the idea of GRB origin in the explosion of the very
bright peculiar supernova SN Ic (hypernova) became very popular.
The features of light curves of optical afterglows and “red bumps”
associated with their nonmonotonic behavior were interpreted as
the accompanying supernovae, which would otherwise remain
unnoticed because of the very large distances to a host galaxy.

\begin{figure}
\centerline{\psfig{file=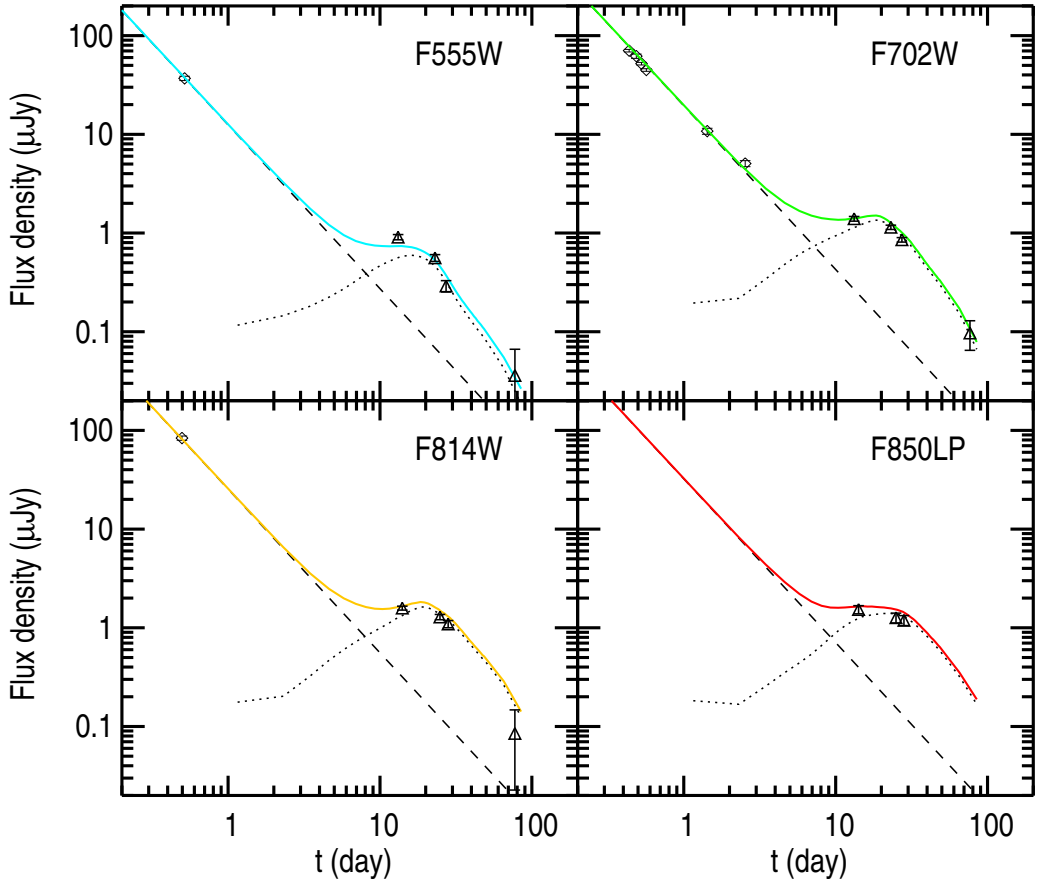,width=5.5in,angle=0}} \caption{(a)
Light curves of afterglow and (b) “red bumps” observable at an
intermediate time in GRB 011121. The triangles correspond to the
photometry with the Hubble space telescope (HST) with the F555W,
F702W, F814W, and F850LP filters (everything with corrections on
an estimate of contribution from the host galaxy), and the rhombs
represent the earth observations. The dashed line describes an
assumed behavior of GRB optical afterglow, and the dotted line
sets an expected flux from a sample supernova for the GRB 011121
redshift, which approximately corresponds to the observational
data with taking into account the 55\% absorption. The solid line
corresponds to the sum of radiations from SN and GRB afterglow
[27].} \label{sn02}
\end{figure}
Another piece of evidence in favor of the GRB–SN relation is
obtained from the detailed photometric and spectroscopic
investigation of the bright optical afterglow of GRB 030329. At a
certain stage, it was noted [100, 43] that the spectra of
afterglow and peculiar supernova SN 1998bw are similar; this
allowed the authors to declare the discovery of SN~2003dh
associated with this GRB.

\begin{figure}
\centerline{\psfig{file=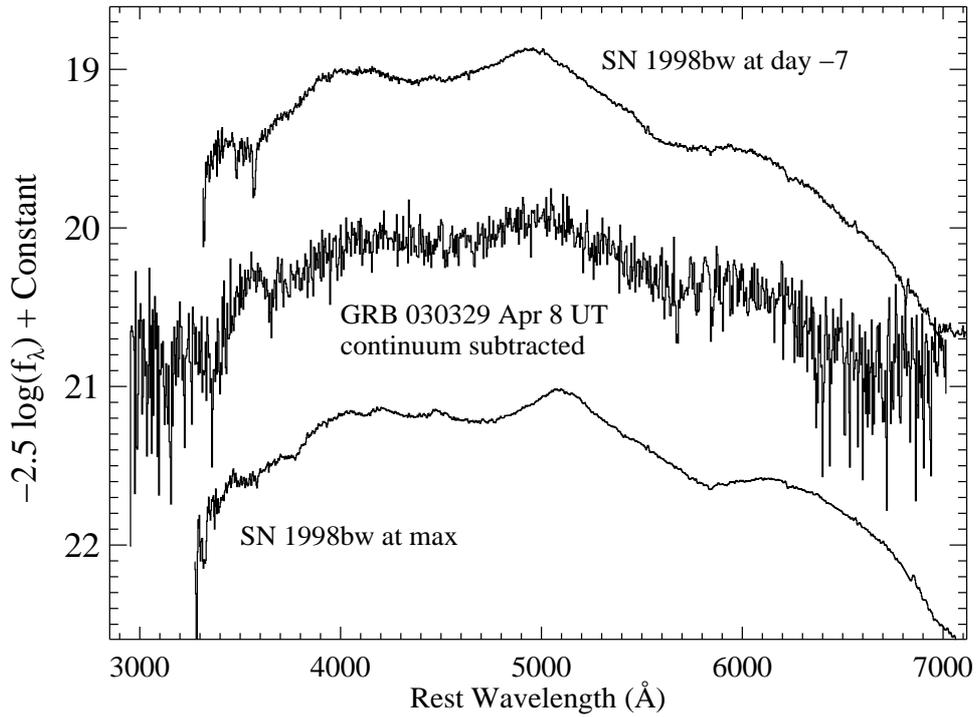,width=5.5in,angle=90}}
\caption{Spectrum of SN 2003dh obtained on April 8 after
subtracting of the smoothed spectrum of April 1. This residual
spectrum shows wide bumps in the regions of approximately 5000\AA
and 4200 \AA (in the rest frame), which are similar to those
observed in the spectrum of a peculiar SN 1998bw Ic type within
one week before the brightness peak [78]. The coincidence is not
so good compared with the spectrum of SN 1998bw in the brightness
peak especially for the red spectral edge [100].}. \label{stanek}
\end{figure}

\begin{figure}
\centerline{\psfig{file=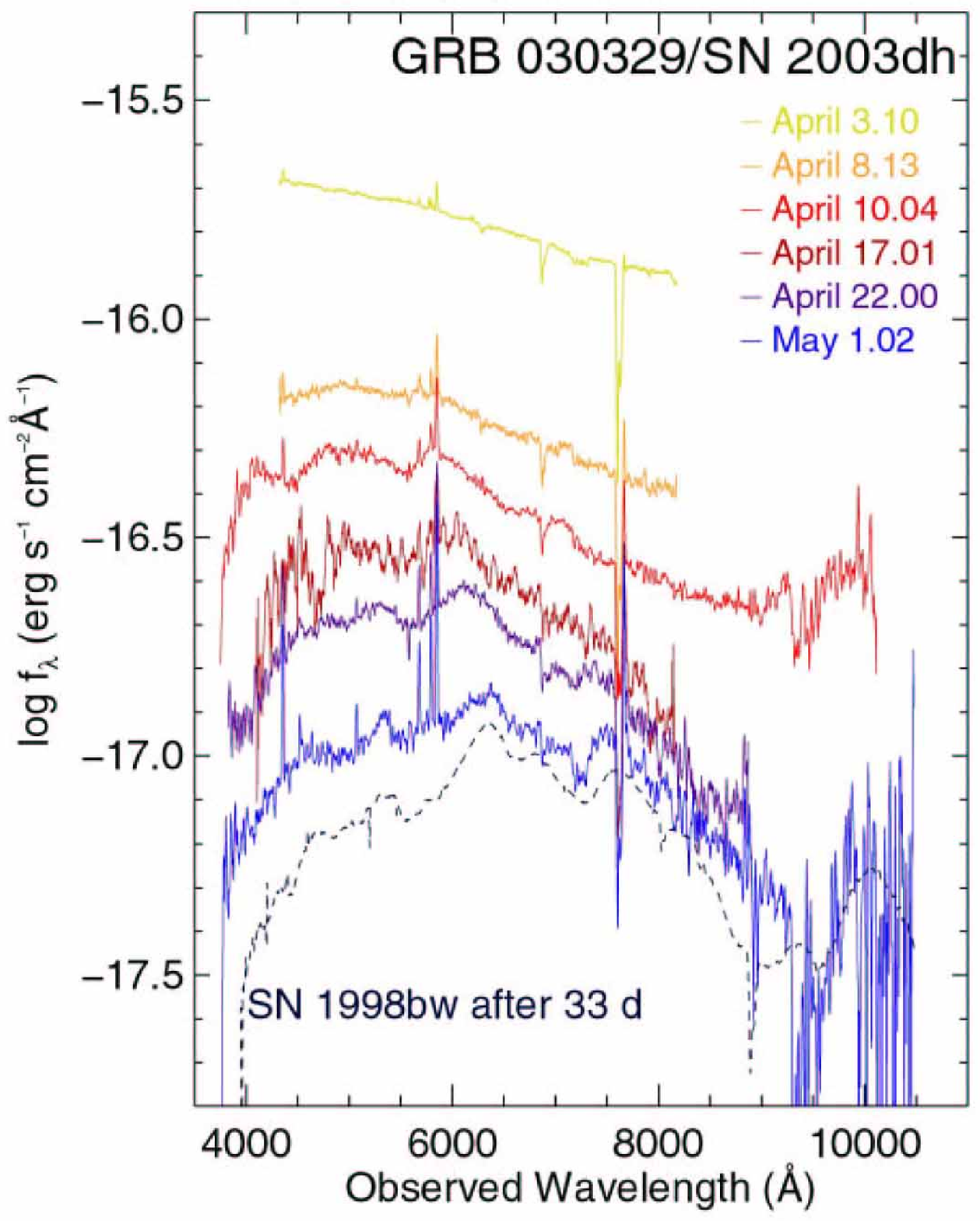,width=5.5in,angle=0}} \caption{
Evolution of the spectrum of total optical radiation for the GRB
030329 afterglow, the SN~2003dh supernova related to it, and their
parent galaxy [43].} \label{hj1}
\end{figure}
We give the comparison between the spectra of SN 2003dh and SN
1998bw in Figs. 17, 18, and 19. In Fig.~18, the upper spectrum is
very well approximated by a power law, which is usually observed
in the spectra of afterglows. The intermediate spectrum shows an
appreciable difference from the power law similar to that observed
in SN 1998bw at the same phase. At the lower spectrum, where SN
2003dh dominates, the supernova features are clearly seen. For
comparison, we show the spectrum of SN 1998bw at the 33d day after
the flare (the dashed line) shifted according to the GRB 030329
redshift. All the SN 2003dh spectra are presented in observed
wavelengths without taking into account corrections on redshift.
The lines of absorption in the atmosphere were also omitted. In
the range of wavelengths longer than 9000~\AA, the spectrum is
strongly polluted with lines of the night airglow, but the wide
detail around 10 000~\AA can be associated with a supernova. At
all stages, the emission lines [O~II]$\lambda$3727, H$_{\beta}$,
[O III]$\lambda$4959 and $\lambda$5007, and H$_{\alpha}$
 are identified as the lines
most likely belonging the host galaxy. At the last stage (May 1),
the following lines also were identified: [Ne III]$\lambda$3869,
H$_\delta$, H$\gamma$, [N II]$\lambda$6583, and the superimposed
lines  He~I$\lambda$3889 + H8 и [Ne III]$\lambda$3968 +
H$_{\varepsilon}$. Strong Balmer lines point to the fact that the
absorption in the host galaxy is weak. The estimate of metallicity
based on fluxes in the lines [O II], [O III], and H$_{\beta}$
gives [O/H] = –1.0. The estimate of a star-formation rate as 0.2
M$_{\odot}$ yr$-1$ is obtained based on the lines [O II] or
H$_\alpha$ Taking into account the 3$\sigma$ upper limit R $>$
22.5$^m$ obtained for the emission of the host galaxy when
studying archive data, it was concluded that equivalent widths of
emission lines are very large. Thus, the host galaxy is the dwarf
with a low metallicity and an active star formation that
qualitatively coincides with the host galaxy of GRB 980425 and SN
1998bw.

\begin{figure}
\centerline{\psfig{file=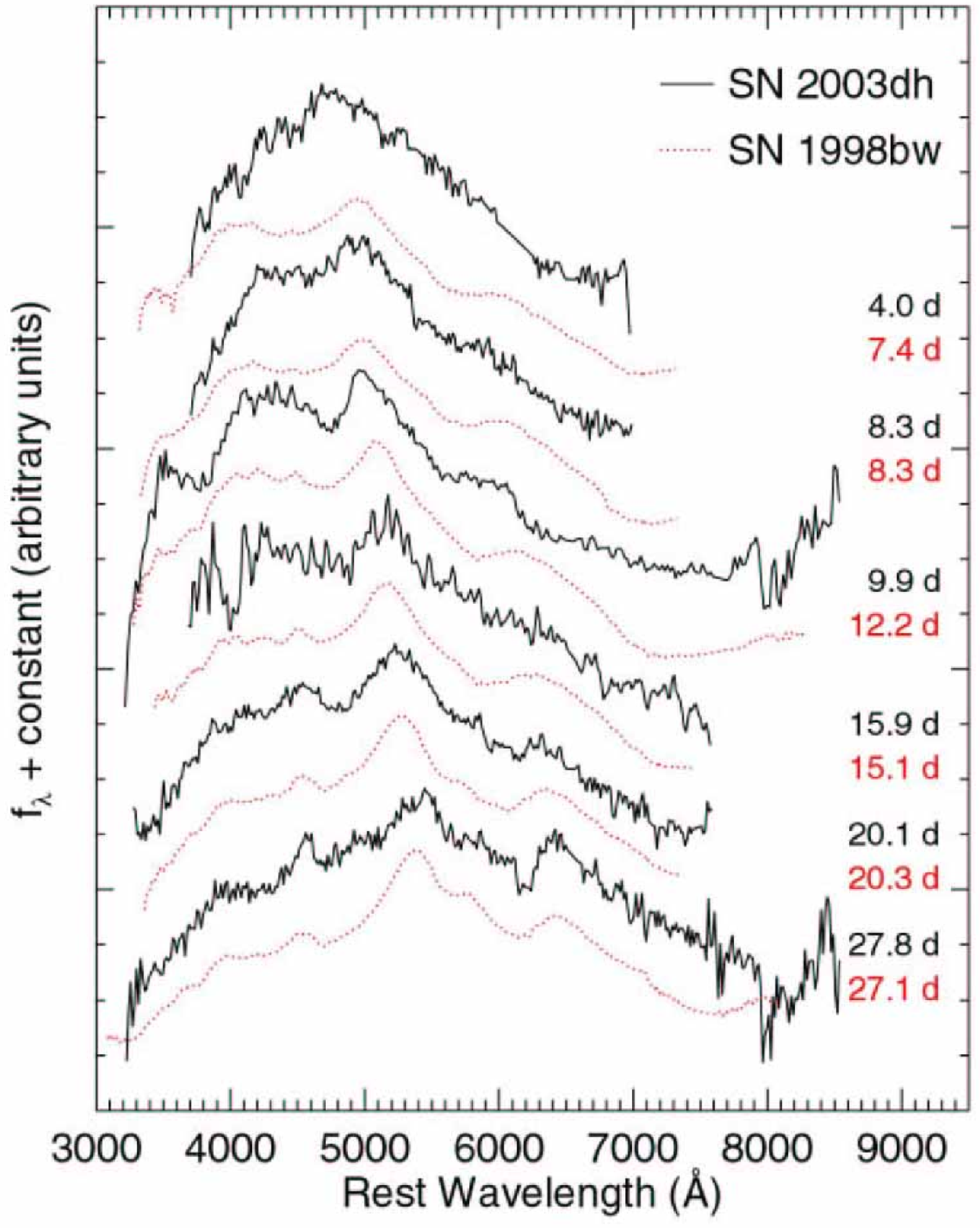,width=5.5in,angle=0}}
\caption{Comparison between the evolution of SN 2003dh and SN
1998bw spectra [43].} \label{hj2}
\end{figure}

In Fig. 19, the solid lines set the SN 2003dh spectra obtained as
a result of the processing described in [43]. The dashed lines set
the SN 1998bw spectra observed at similar epochs. We give all the
spectra in the restframe wavelengths and marked by indicating days
from the GRB-flare moment with taking into account the
cosmological time extension $(1 + z)$. The relative values of
spectra along vertical are arbitrary. The SN 2003dh spectrum was
divided into 15-\AA regions, the lines of the host-galaxy
radiation were removed, and an interpolation was made where it was
necessary in the region of strong night-sky absorption lines at
6800~\AA, 7200–7400~\AA, and 7600~\AA. The strong absorption
"edge" at 6150~\AA (observed at 7200~\AA) is likely associated
with the night sky absorption. When processing the spectra, it was
assumed that these are only the host-galaxy emission lines that
were visible and subtracted. The spectrum was fit in the form of
the sum of a power part ($f_{\lambda} \sim \lambda_{-(\beta+2)}$)
and the scaled spectrum of the sample SN 1998bw. Using the least
square technique, three parameters have been obtained: the
spectral index $\beta$, the afterglow amplitude, and the supernova
amplitude. In most cases, the best index was obtained equal to
$\beta \sim -1.2 \pm 0.05$, which was used everywhere; however,
this index and the sample spectrum shape only slightly affected
the estimate of contribution of supernova in the spectrum shape.
The spectral-peak wavelength redshifts in time for both supernova,
and the average rate of such a shift amounts to $\sim 25$\AA per
day for SN~2003dh that is similar to the early evolution of the SN
1998bw spectrum. In [43], an increase in opacity on absorption is
assumed as the cause of such a shift for the motion in bluer
region from 4900~\AA (rest-frame wavelength).

Figure 20 shows the earliest optical-afterglow spectra of GRB
030329 observed in [99]. From the comparison between Figs. 18 and
20, it follows that, at early moments, for example, within 10–12 h
after the GRB, the spectral shape was far from that described by
the power law to which it evolved within several days to April 3.
In Fig. 20, it could be interpreted as the spectrum of an initial
strong outburst of SN 2003dh similar to a short initial outburst
observed on the light curve for SN 1993J shown in Fig. 21. The
model of such an outburst was calculated in [95].

Despite such striking coincidences, the relation between GRB and
SN still cannot be considered as firmly established. There are
spectral distinctions, time scales differ from standard values,
and the spectra of many GRB afterglows involve no indications of
the presence of supernovae. In addition, in a unique case when the
coinciding GRB and SN were discovered independently, a quite
ordinary observable GRB 980425 appeared together with SN 1998bw
proved to be very abnormal, being three orders of magnitude weaker
than other cosmological GRBs with known redshifts of lines in
afterglows (see Table 2).

\begin{figure}
\centerline{\psfig{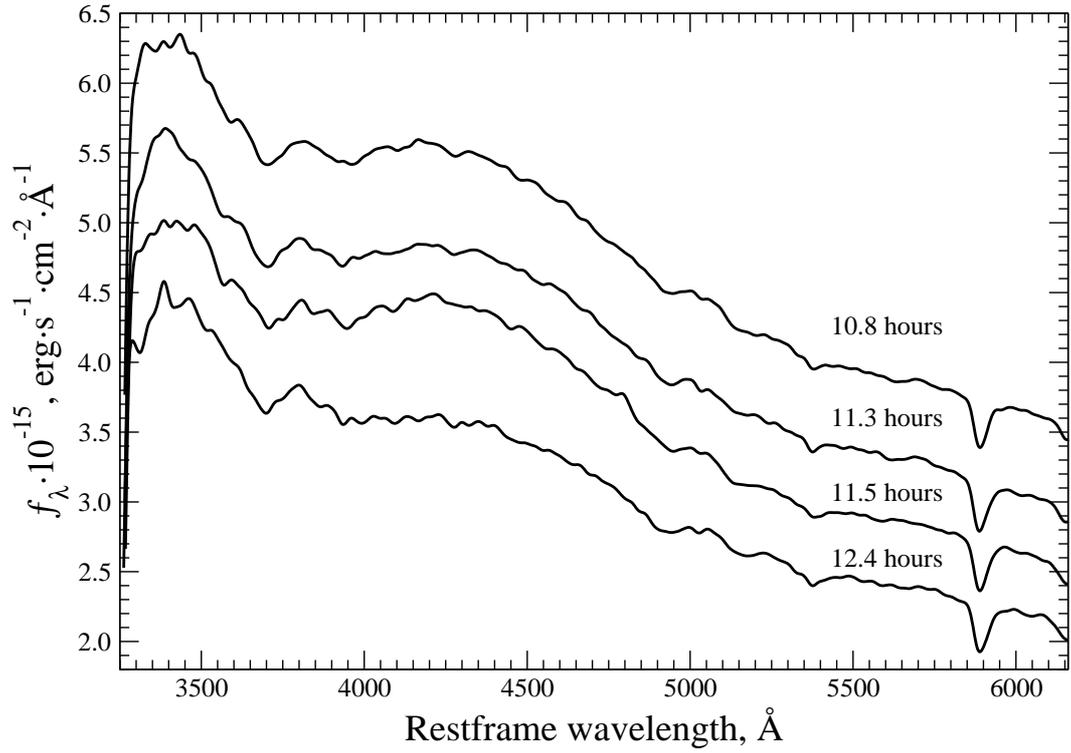}}
\caption{Smoothed spectra of GRB 030329 optical afterglow in
rest-frame wavelengths obtained with the spectral resolution of 12
A by the 6-m telescope SAO. The spectra obtained at the time
moments of 11.5, 11.3, and 10.8 h after the GRB are shifted along
the $f_\lambda$ vertical scale by $+0.2\cdot 10^{-15}$, $+0.6\cdot
10^{-15}$ and $+1.2\cdot 10^{-15}$, respectively, [99].}
\label{sokf2}
\end{figure}

\begin{figure}
\centerline{\psfig{file=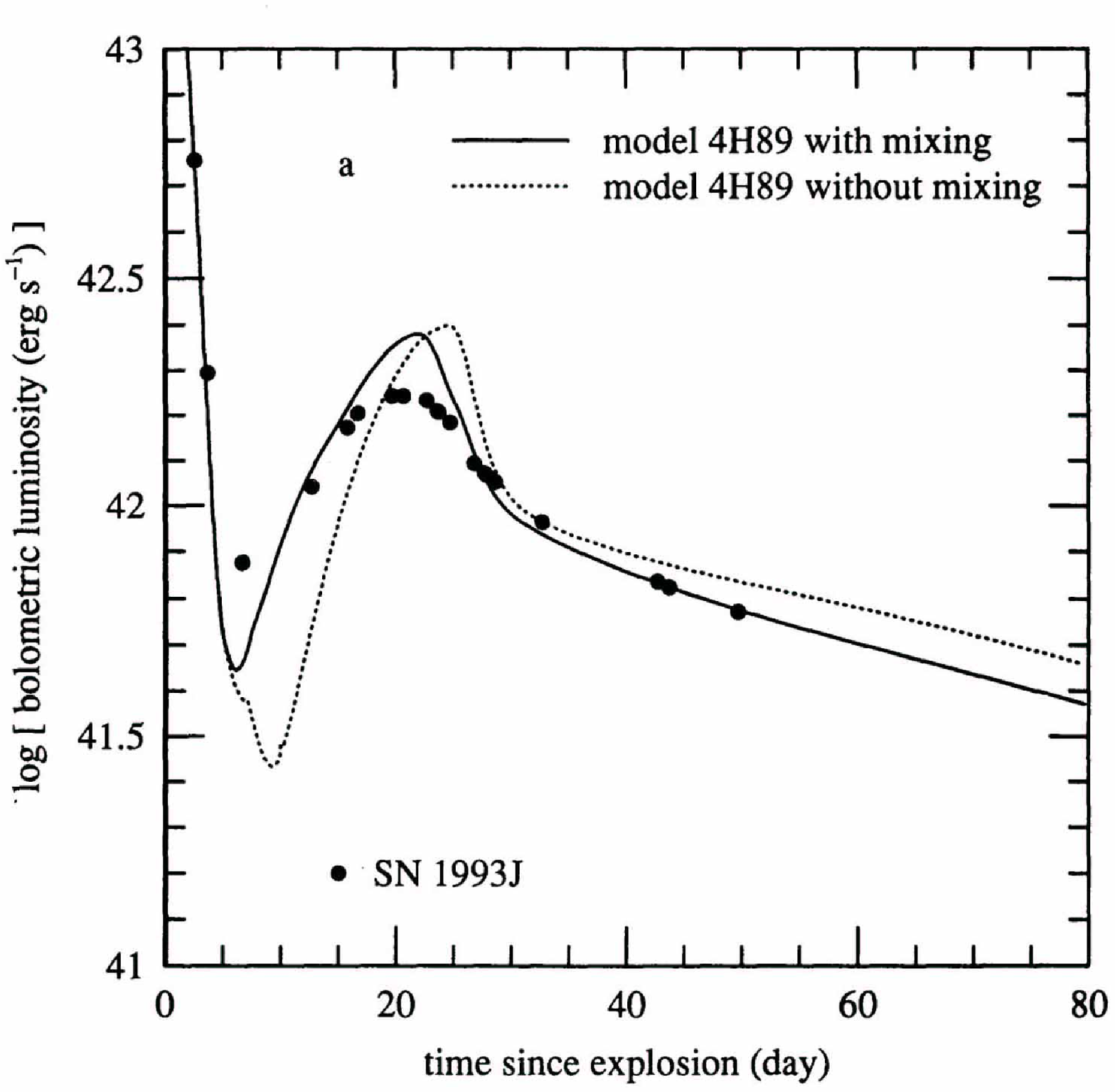,width=5.5in,angle=0}}
\caption{Light curves calculated using the model with a shell mass
of 0.89$M_{\odot}$. Bolometric luminosity measurements for SN
1993J are given by closed circles [95].} \label{sn1993}
\end{figure}

\section{Polarization of gamma rays in GRB}

The authors of [30] reported on the discovery of the linear
polarization of . rays from GRB 021206. The authors interpreted
this fact as a synchrotron radiation of relativistic electrons in
a strong magnetic field. Measurements were carried out at the
Reuven Ramaty High Energy Solar Spectroscopic Imager (RHESSI). The
RHESSI has a set of nine coaxial germanium detectors of a large
fiducial volume (300 cm3) with a high spectral resolution for
studying x and $\gamma$ rays from the Sun in the energy range of 3
keV–17 MeV. The RHESSI has a high angular resolution (2$''$) with
a field of view of $\sim$1$^{\circ}$ of its optics; however, flat
focal detectors are unscreened and cover the entire sky. Because
of this fact, the RHESSI very frequently detects GRBs in the focal
plane of detectors, although the probability to fall into the
image-construction field is low. These observations allow the
authors to obtain highly resolved spectra, while detection times
and energies of individual photons are potentially suitable for
measuring the polarization. The RHESSI is not optimized for use as
a $\gamma$-polarimeter, but some features of its design make it
the most sensitive of modern devices for astrophysical
measurements of the $\gamma$-ray polarization [30].

For the soft $\gamma$ rays with $\sim$0.15–-2.0 MeV, the basic
mechanism of interaction of photons with RHESSI detectors is the
Compton scattering. A small fraction of incident photons undergoes
the single scattering in one of detectors before being scattered
or photoabsorbed in another detector. The probability of this
sequence of events depends on the polarization of the incident
$\gamma$ ray. The linearly polarized $\gamma$-rays scatter
predominantly in the direction perpendicular to their polarization
vector. In the RHESSI, this property of scattering can be used for
measuring the polarization of astrophysical sources. The device
sensitivity to the polarization depends on an effective scattering
area and an average value of polarimetric modulation depth
$\mu(\theta,E)$, which is equal to a maximal spread in the
probability of azimuthal scattering of polarized photons. This
depth is equal to $\mu = (d\sigma_{\perp} -
d\sigma_{\parallel})/(d\sigma_{\perp} + d\sigma_{\parallel})$,
where $d\sigma_{\perp}$, and $d\sigma_{\parallel}$ are the
Klein–Nishina differential cross sections for the Compton
scattering in perpendicular and parallel directions relative to
the polarization direction, respectively. It is a function of the
incident-photon energy $E_{\gamma}$ and the Compton scattering
angle $\theta$ between the direction of incident and scattered
photons. For a source with the count number $S$ and the degree of
polarization $\Pi_{s}$, the distribution of expected azimuthal
scattering angles is equal to $dS/d\phi =
(S/2\pi)[1-\mu_{m}\Pi_{s}\cos(2(\phi-\eta))]$ , where  $\phi$ is
the azimuthal scattering angle, $\eta$ is the polarization- vector
direction, and $\mu_{m}$ is the average value of polarimetric
modulation depth of the device. Although the RHESSI has a small
effective area ($\sim$20\,cm$^2$) for the events in which the
scattering between detectors takes place, it has a very large
modulation in the region of 0.15–2.0 MeV obtained in the Monte
Carlo calculations. According to [30], the results of processing
of observational data shown in Fig. 22 convincingly point to the
discovery of a strong polarization for the GRB direct $\gamma$
rays. The authors of [30] assumed that this result specifies the
important role of the magnetic field in the GRB-explosion
mechanism. It agrees with the GRB model in the form of a fireball
inflated by the magnetic field. This field could be generated due
to the rotational energy of an accretion disk differentially
rotating around a central compact object, or the rotation energy
of the Kerr black hole pierced by magnetic field lines, or the
rotational energy of a strongly magnetized neutron star [105].
Another possible model is the generation of a large-scale magnetic
field in a flux beyond the shock-wave front.

\begin{figure}
\centerline{\psfig{file=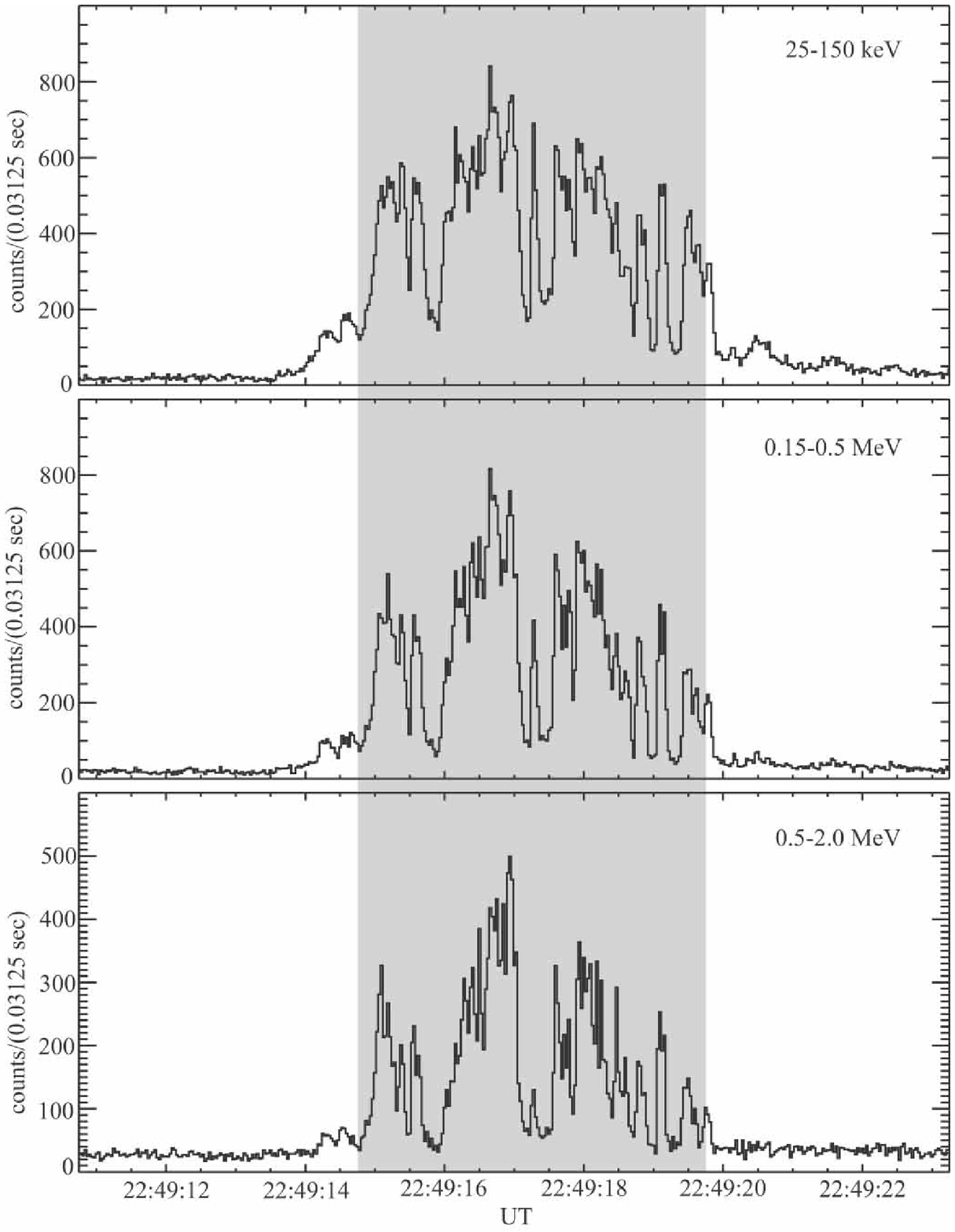,width=7cm,angle=0}
            \psfig{file=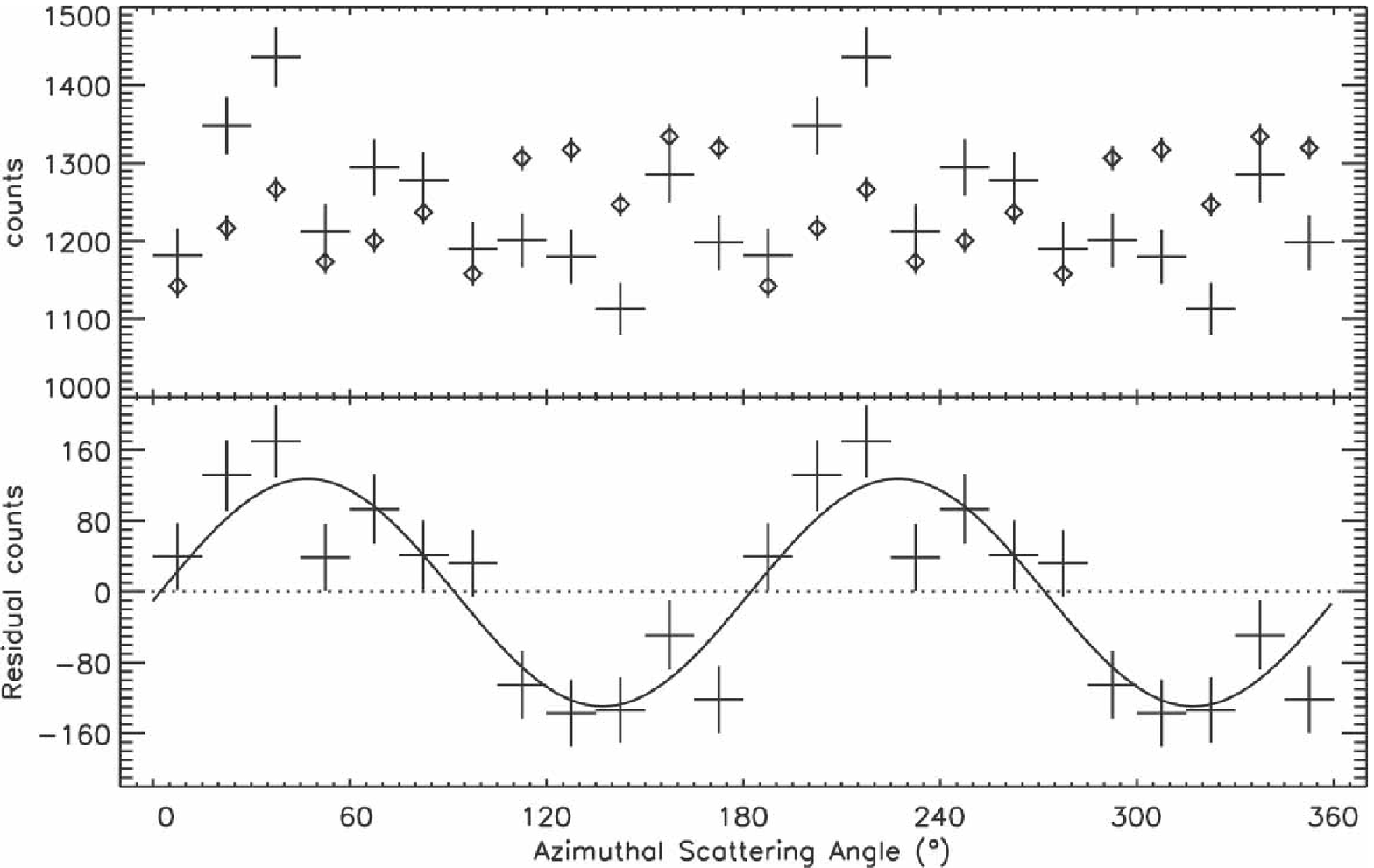,width=7cm,angle=0}}
\caption{ On the left: RHESSI light curves (total number of
counts) in three energy intervals for GRB 21206. This GRB is
localized in the sky at 18$^{\circ}$ from the Sun, ruling out the
possible search for the optical afterglow; however, its
brightness, duration, and proximity to the RHESSI rotation axis
make it an ideal candidate for searching the polarization. The
shaded region shows the signal with a 5-s averaging time for the
analysis of polarization. On the right: the azimuthal distribution
of scatterings for the RHESSI data corrected for the satellite
rotation. The angular interval of 0$^{\circ}$--180$^{\circ}$ is
divided into subintervals of 15$^{\circ}$ in which the counts
represented here were gathered twice for the sake of clarity. The
upper graph shows the unprocessed measured distribution (crosses)
together with the result of the Monte Carlo simulation for a
simulated distribution of an unpolarized source with a constant
flux (rhombs). The lower graph shows the RHESSI data after
subtracting the simulated distribution. The obtained residual
distribution is incompatible with the unpolarized source (the
dashed line) at the confidence level higher than $5.7\sigma$. The
solid line shows the result of the best fit of the modulation
curve corresponding to the degree of polarization of $(80\pm20)$\%
[30].} \label{gpolar1}
\end{figure}
Data [30] were analyzed again in [90], where it was concluded that
they give no hint of the detection of a polarized signal (see Fig.
23). First, it was found that the number of polarization events
used for the measurement, where the photon passed from one
detector to another as a result of scattering and was detected in
both, was approximately ten times lower than that estimated in
[30] $830 \pm 150$ instead of $9840 \pm 96$). As a result, the
signal-to-noise ratio proves to be too low for detecting a
possible polarization even at the level of 100\%. In [90], the
authors used another method of data processing for detecting the
polarization effect and revealed no indications of the
polarization in the GRB 021206 radiation. It was shown that the
detected signal can correspond to an unpolarized source, and the
polarization observed in [30] is associated with neglecting a
systematic error in the “zero light curve” used in processing.
Because of the low signal-to-noise ratio in the RHESSI data, the
analysis made in [90] by a method which can be employed only for
the analysis of Poisson noise, demonstrated that the data could be
consistent with any polarization up to 100\%. This points to the
fact that it is impossible to obtain estimates of polarization of
the GRB 021206 radiation from the RHESSI observations.

\begin{figure}
\centerline{\psfig{file=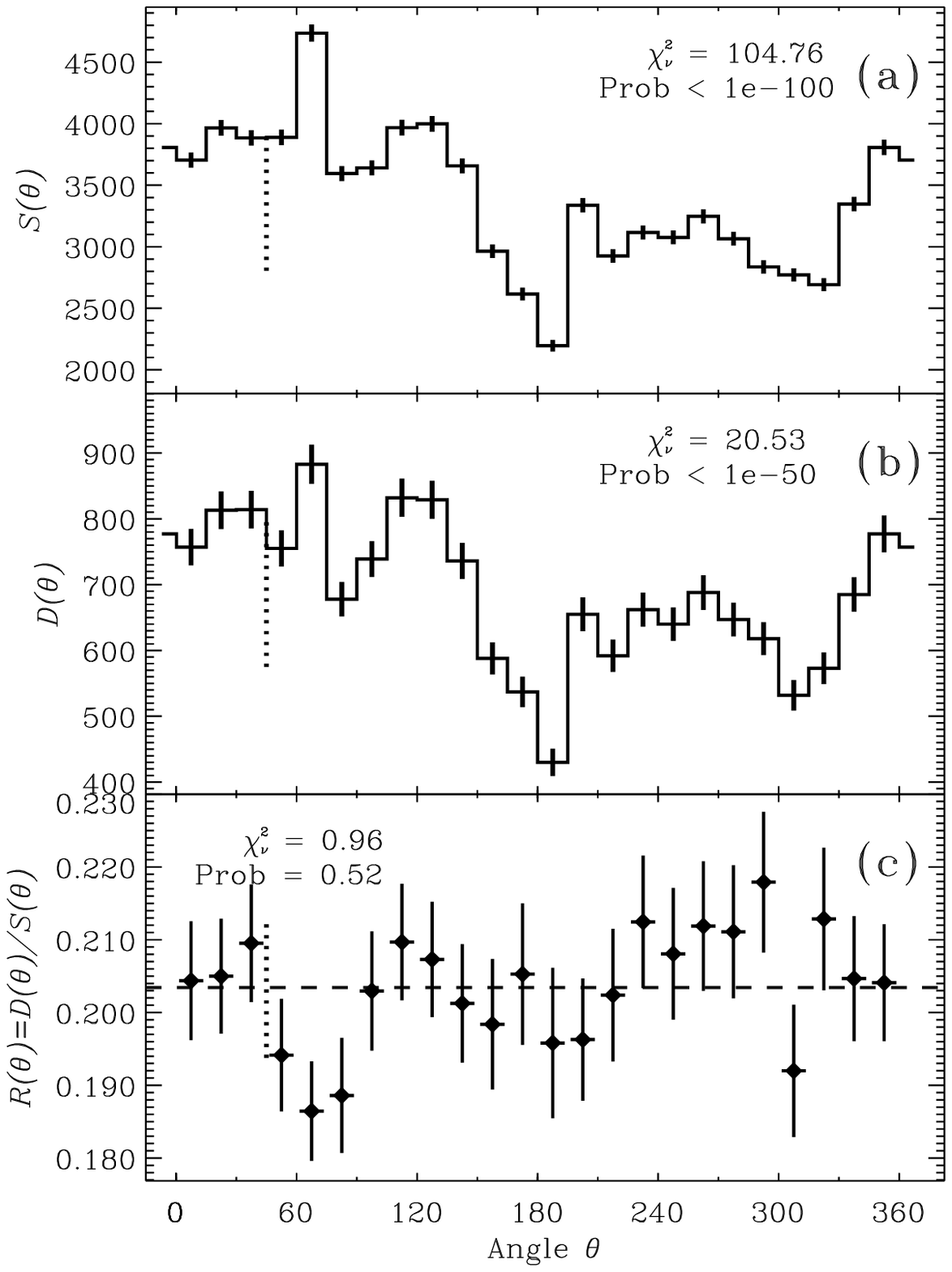,width=5.5cm,angle=0}
 \psfig{file=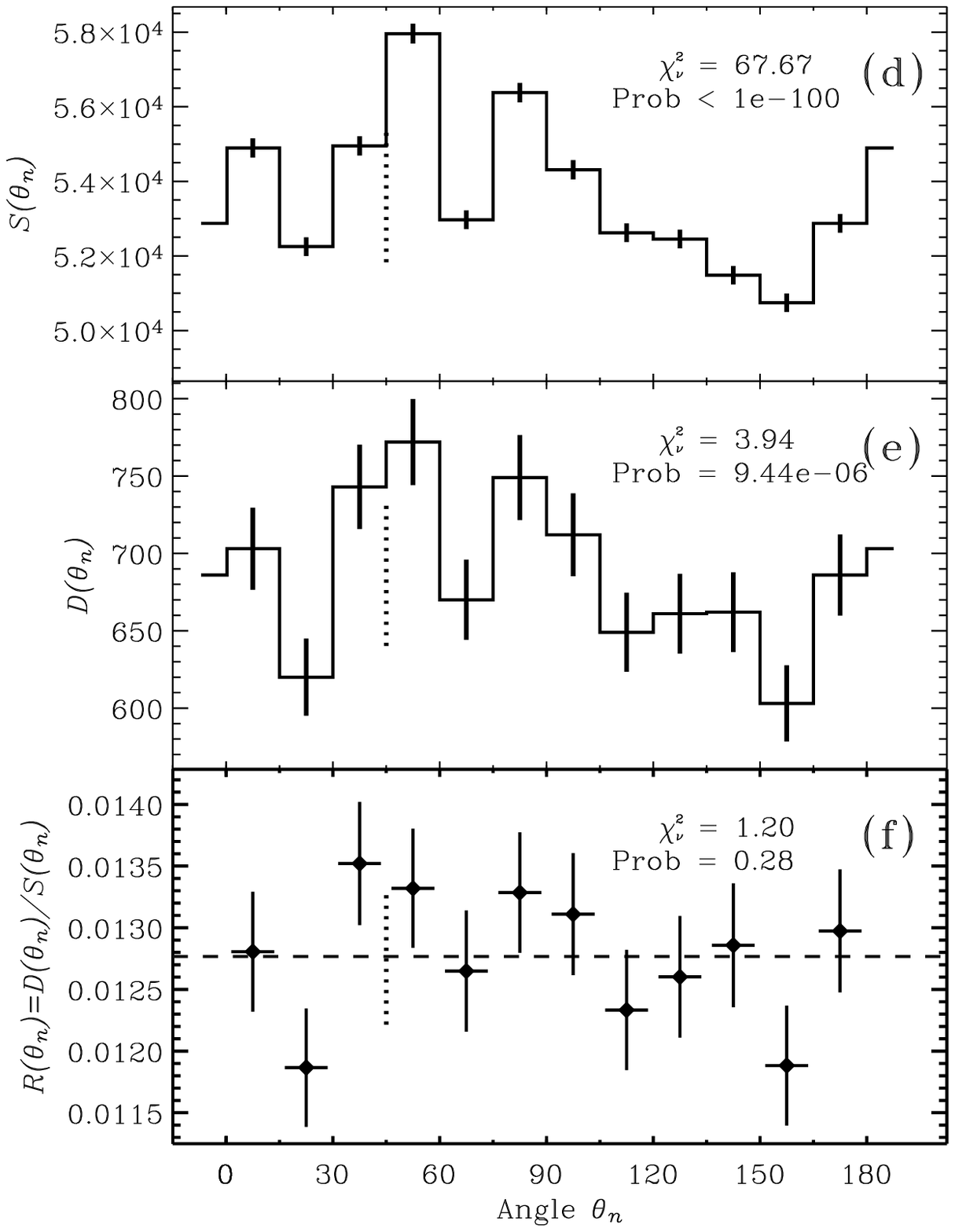,width=5.5cm,angle=0}}
          \caption{Here, $\theta=0 $ и $\theta_n$=0 correspond to the
          direction towards the celestial north with an increase
          in the angle when rotating
the detector base from the north to the east (clockwise facing the
Sun). The GRB 021206 position in all the figures is shown by the
vertical dotted line. {\bf (a)} The counting rate $S$ of solitary
events as a function of $\theta$, where $\theta$ is the angle
between the celestial north and the center of the detector
detecting a photon in the interval of 0-360$^{\circ}$. $S(\theta$)
is incompatible with a constant value, $\xi_{\nu}^2$ is indicated;
the observation probability of such a variability is lower than
$10^{-100}$. {\bf (b)} The counting rate$D$ of double events as a
function of $\theta$; here, both counts are indicated for each
event located in the corresponding intervals of $\theta$. The
observed variability is incompatible with the constancy of
$D(\theta$),  $\xi_{\nu}^2$ is indicated; the observation
probability of such a variability is lower than $10^{-50}$. {\bf
(c)} The ratio $S/D$ is compatible with constant with the
probability equal to 0.54. {\bf (d)} The counting rate
$S$($\theta_n$) of solitary events for the polarization analysis
in the interval of 0-180$^{\circ}$. The counting rate is
incompatible with constant. {\bf (e)} The counting rate
$D$($\theta_n$) of double events for the polarization analysis in
the interval of 0-180$^{\circ}$ is well compatible with the
corresponding graph from [30] without subtracting the Monte Carlo
“zero” light curve. An essential variability on $\theta_n$ is
seen. {\bf (f)} The ratio $R=D(\theta_n)/S(\theta_n)$. The
polarized photons would scatter in a sky predominate direction
with a higher ratio of double-to-solitary counts than in the
perpendicular direction leading to a variability in $R(\theta_n)$.
From the observations, it follows that $R$($\theta_n$) is
compatible with constant; i.e., no indications of polarization
were found [90].} \label{gpolar2}
\end{figure}
In reply to [90], the authors published a new paper rejecting this
criticism. It was noted that [90]  "raises many important problems
associated with processing the observations, which will be
considered in a separate paper being prepared for publication.
However, it should be noted that the limit for the detected degree
of polarization obtained in [90] with the use of the new method of
processing developed there is strongly overestimated. Although we
would like to learn in more detail the new statistical technique
proposed in [90] (especially with the substantiation of the
assumption that the obtained degree of polarization is independent
of the instrumental response), we come to an inevitable conclusion
that there is a serious gap in their statistical method. In
addition, the authors of [90] declared that their analysis is
insensitive to an arbitrary level of polarization. Hence, it does
not contradict that level of polarization, which was presented in
our original paper." The investigation performed in [113]
confirmed the conclusion of [90] that it is impossible to
determine the degree of polarization of GRB radiation from the
available data. It is possible that the further processing method
development and new observations are necessary for overcoming this
contradiction; however, the majority is on the side of skeptics.

Nevertheless, even without the assurance in reliability of
measurements of the polarization for the direct GRB radiation,
certain theoretical models were developed for explaining this
phenomenon with the use of both synchrotron radiation [38] and
inverse Compton scattering [57]. In [38], the degree of
polarization $P\sim 43-61\%$ was found for an ordered transverse
magnetic field $B_{\rm ord}$. At the same time, the field
$B_\perp$ generated on the shock-wave front, which is chaotic but
located completely in the shock-wave plane, can provide the degree
of polarization up to $P\leq 38-54\%$ in a solitary pulse on the
GRB light curve. However, it is expected that the polarization in
the integrated radiation of several pulses measured for GRB 021206
decreases approximately by half. The magnetic field $B_\parallel$
normal to the shock-wave front can provide the degree of
polarization  $P\sim 35-62\%$  for the total radiation of many
pulses. However, the measurements of the polarization in GRB
afterglows assume a more isotropic configuration of the field
generated by the shock wave that should reduce $P$ $\sim 2–3$
times. Therefore, the ordered field $B_{\rm ord}$ arising in a
source can explain the observed polarization most naturally, while
$B_\parallel$ is less probable, and $B_\perp$ is least expected.

In [57], the inverse Compton scattered radiation of a fireball was
considered as the directed jet with an opening angle comparable to
or larger than the collimation angle associated with the
relativistic contraction of the beam. The degree of linear
polarization as a function of the jet-opening angle and the
direction to observer were numerically calculated for the Compton
upward scattering along the flux for a large-scale motion (see
Fig. 24). Note that the gamma factor of the large-scale motion
cannot be large due to the absence of apparent inverse dependence
of the GRB duration on its power, which is inherent to the model
with a large-scale relativistic motion (see section 4.1).

\begin{figure}
\centerline{\psfig{file=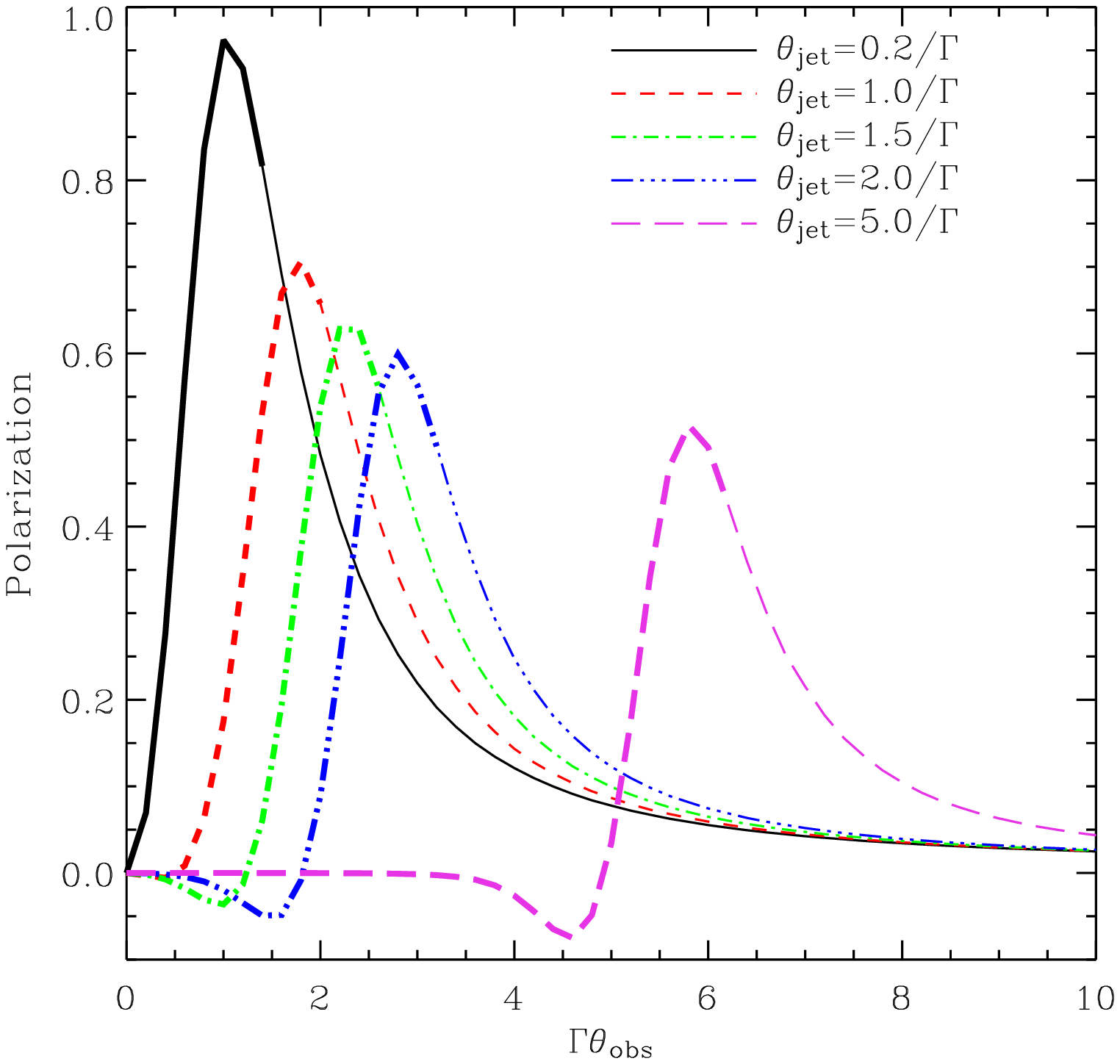,width=7cm,angle=0}
            \psfig{file=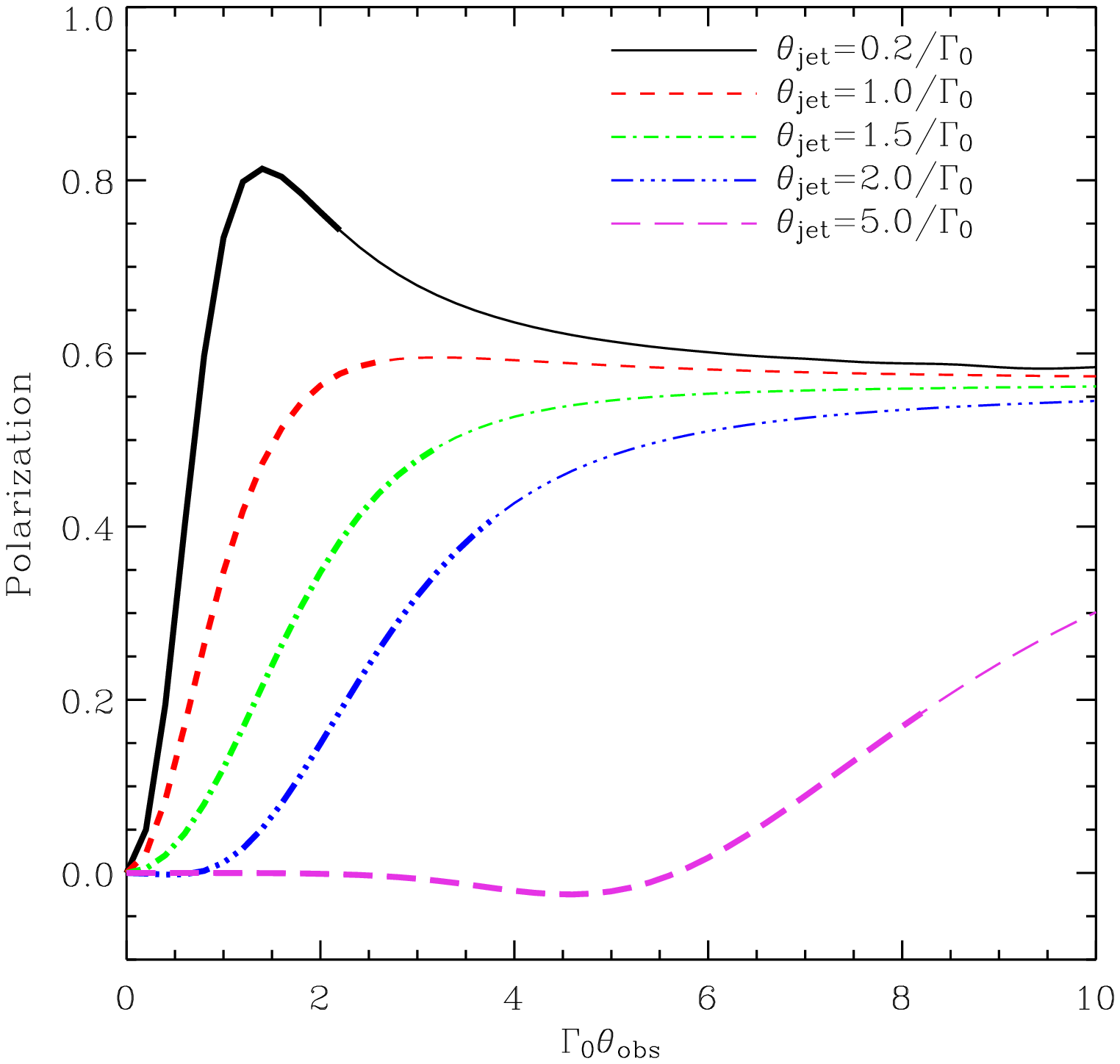,width=7cm,angle=0}}
\caption{On the left: the degree of polarization as a function of
the observation angle $\theta_{obs}$ in units of $1/\Gamma$ for a
homogeneous jet with sharp boundaries. Various lines represent the
polarization for jets with various opening angles. A bolder line
designates the region where the scattering efficiency is higher
than 2.5\%. On the right: the same picture for a jet with the
Gaussian distribution. Because the Lorentz factor is variable in
this jet, the observation angle $\theta_{obs}$ is indicated along
the axis $x$ in units of $1/\Gamma_0$, i.e., in units of the
Lorentz factor along the jet axis [57].} \label{laza}
\end{figure}

\section{Observation of x-ray echo}

The observations of GRB 031203 at the Newton x-ray observatory
(XMM-Newton) began on December 4, 2003 at 04:09:29 UT and lasted
58 211 s. This GRB was detected for the first time by the IBIS
device on the satellite INTEGRAL 2003-12-03 at 22:01:28 UT. The
analysis of the first observations by the Newton telescope found a
diffusive x-ray halo around the afterglow site. All three chambers
of the EPIC device observed this halo, and it is not associated
with the scattering of optical or $x$-ray photons inside the
device. The halo shape was almost a perfect circle, the radius of
which increased in time. This recalls the behavior of the light
echo going from x-ray photons scattered by the dust of a cloud
located at a distance of about 700 pc from the observer [59]. GRB
031203 has galactic coordinates l = 255.74$^\circ$ and b =
–4.80$^\circ$ in the direction of which the Gum Nebula, other
nebulas, and infrared sources are located. The distance to the
scattering cloud points to its location inside our Galaxy. The
$x$-ray spectrum of GRB 31203 is well approximated by the power
law with a photon index of $\sim 1.7$. As expected, the scattered
x-rays have a softer spectrum with a photon index of $\sim 3$
[59]. The subsequent analysis shows [107] that the halo appears in
the form of concentric circular rings with the center at the GRB
localization (see Fig. 25). The radii of these rings grow in time
as $\sim t^{1/2}$, which is compatible with a small angle of
scattering of $x$-ray photons on a large dust column along the
direction of observation of the cosmological GRB (Fig. 26). The
appearance of two rings is associated with two different dust
layers in the Galaxy located at distances of 880 and 1390 pc,
which agrees with known features of the Galaxy structure. The halo
brightness points to an initial momentum of the soft $x$ rays
emitted simultaneously with the observed GRB.

We note that the same $\sim t^{1/2}$ law for an increase of the
ring radius in time takes place for an arbitrary variant when the
ring radius $(x)$ is much less than the distance from a dust cloud
to both the observer $(d_1)$ and the GRB source $(d_2)$. Always
assuming that $x\ll d_1$, we obtain

\begin{equation}
\label{xech1} x=d_1\sqrt{2}\Biggl[\frac{ct}{d_1}+\frac{d_2}{d_1}+1
   -\sqrt{\left(\frac{d_2}{d_1}+1\right)^2+2\frac{ct}{d_1}}\Biggr]^{1/2}.
\end{equation}
Here, $d_2$ is arbitrary, and the time is counted from the moment
of detection of the direct GRB $x$-ray ($\Gamma$-ray) pulse. When
the source is inside the cloud ($d_2=0$), we have from Eq. (8)
that

\begin{equation}
\label{xech2} x\approx ct.
\end{equation}
In the case of $d_2 \gg x$, we have

\begin{equation}
\label{xech3} x=\sqrt{2ct
\left(\frac{1}{d_1}+\frac{1}{d_2}\right)^{-1}},
\end{equation}
i.e., the same square-root time dependence. Here, the only
requirement is large distances  $d_1$ and $d_2$ in comparison with
$x$, which can be fulfilled even for the GRBs located at
noncosmological distances. Thus, an increase in the radius $x(t)$
according to the square-root law is compatible with the
cosmological GRB nature, but cannot be considered as proof of this
nature.

\begin{figure}
\centerline{\psfig{file=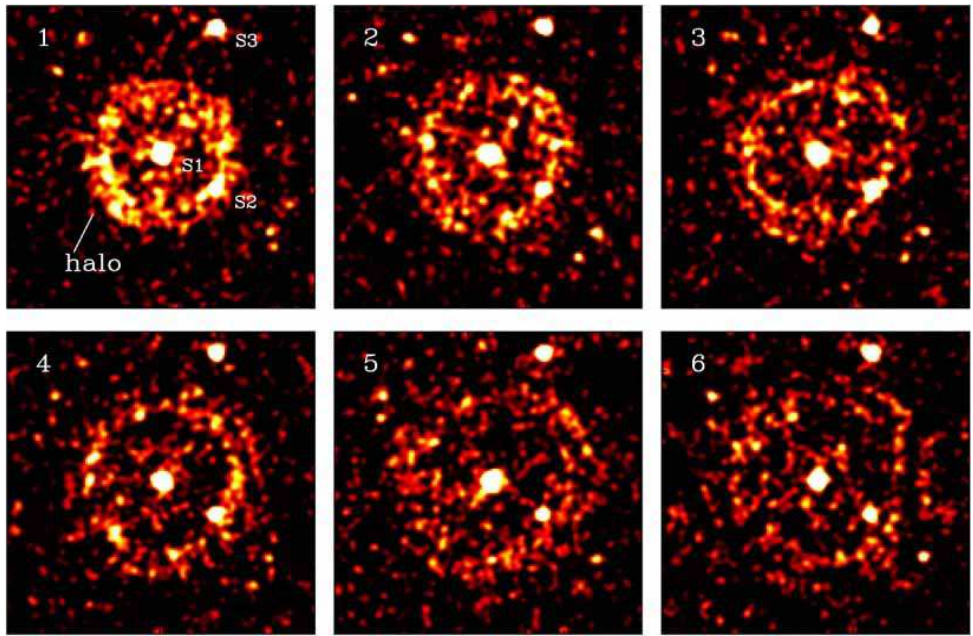,width=5in,angle=-90}
            }
\caption{Series of images of the region in ten angular minutes
around GRB 031203 in the energy region of 0.7–2.5 keV. The
observational data are divided into ten time intervals of 5780 s.
The first six images are smoothed with the use of a Gaussian
nucleus of six angular seconds in size. The three brightest point
sources (S1, S2, and S3) are marked [107].} \label{echo}
\end{figure}

\begin{figure}
\centerline{\psfig{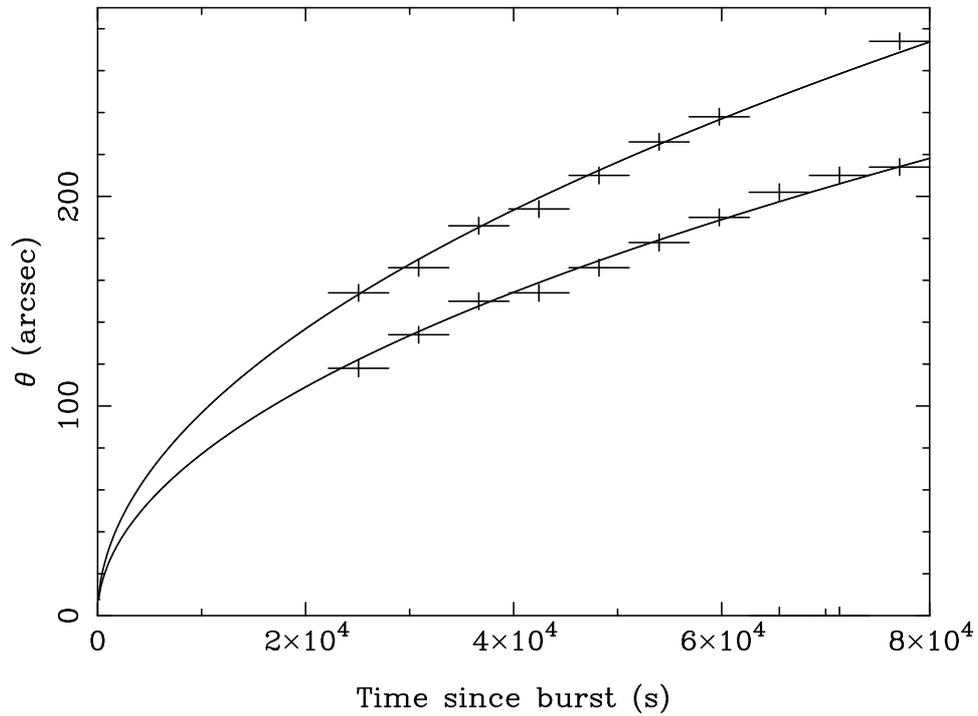}
            }
\caption{Time increase in the radius of two rings around GRB
031203. The radius of each ring was measured from the local peak
along the radius. In both cases, the distribution is well
approximated by the function  $\theta \propto (t-t_0)^{1/2}$. A
segment length used in measurements of profile along radius was
taken as an error [107].} \label{echo1}
\end{figure}

\section{Reaction of a dense molecular
cloud to a nearby cosmological GRB}

\subsection{Optical afterglow
from interstellar-gas reradiation}

As was noted for the first time in [76] (see also [81]), the
GRB-afterglow properties are better explained under the assumption
that GRB sources are located in the star-formation regions with a
high gas and dust density. The interaction of a powerful GRB pulse
with the surrounding gas of a density $n=10^4\,-\,10^5$ см$^{-3}$
results in the occurrence of a special-form optical afterglow the
duration of which can reach ten years. The light curve and the
spectrum of such an afterglow were calculated in [25] (see Fig.
27). It was shown that the duration and spectrum of the GRB
optical afterglow depend on the GRB location with respect to the
dense cloud because of the interaction of $\gamma$ rays with the
dense interstellar medium. In the case of an anisotropic GRB, they
depend also on the orientation of the observer and the cloud with
respect to the GRB axis. During the explosion in a homogeneous
cloud, the optical afterglow can endure for ten years. The GRB
optical afterglow can be distinguished from a supernova radiation
with the same energy yield by the light-curve and spectral
features. The $H_{\alpha}$ and $H_{\beta}$ emission lines are
strongest in the optical afterglow. The discovery of the similar
afterglow would give important information on the GRB-source
properties and the surrounding medium.

\begin{figure}
\centerline {\psfig{file=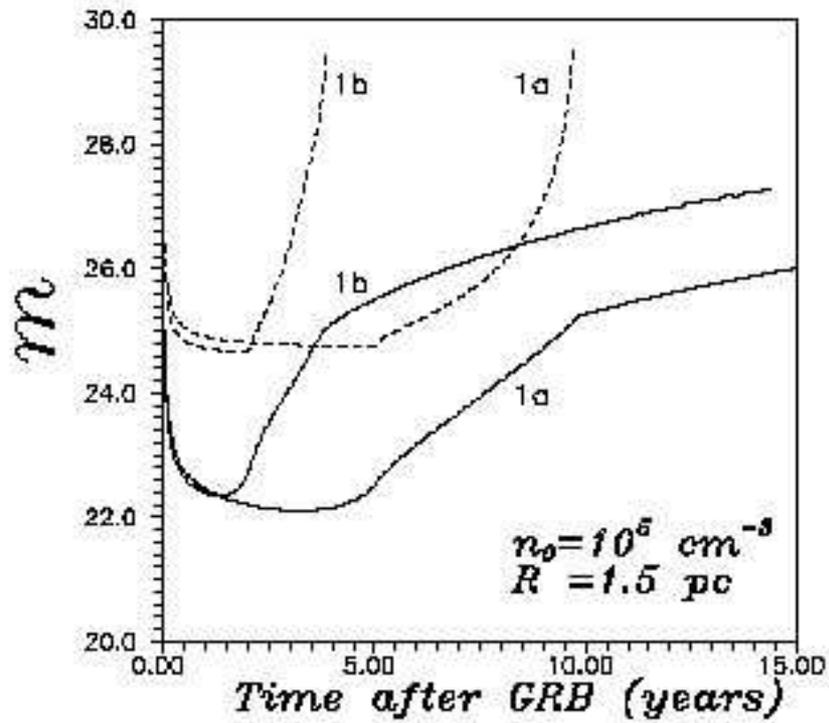,width=12cm,angle=-0}
 }
\caption{27. Light curve of the optical afterglow in stellar
magnitudes. The solid line gives the upper limit, and the dashed
line the lower limit. The origin of coordinates corresponds to the
moment of the GRB flare with an energy $E$ and a total fluence
near the Earth $F_{\rm GRB} = 10^{-4}$ erg cm$^{-2}$: curves 1a
and 1b correspond to the cases of $E = 10^{52}$ erg, $n_0= 10^5$
см$^{-3}$; and $E = 10^{51}$ erg, $n_0= 10^5$ см$^{-3}$,
respectively, [25].} \label{bktim}
\end{figure}
More detailed 2D calculations of the GRB afterglow after the
explosion in the star-formation region nearby a cold molecular
cloud were made in [9, 10]. Optical light curves qualitatively
coincide with those in Fig. 27; however, the large variety of
spectra and light curves takes place depending on initial
distributions of density in cloud, the anisotropy of density
distribution, and GRB radiation. For solving the equations of
hydrodynamics with the radiation heating and cooling in the
complete-transparency approximation, the PPM method was used. As
in [25], the GRB propagation through a cloud was approximated by a
very narrow spherical or directed gamma-photon wave leaving behind
a completely ionized gas heated by the Compton interaction. When
calculating the heating, the Klein–Nishina corrections to the
Thomson cross section [11] were taken into account. The
temperature distribution after an isotropic GRB explosion in the
center of a homogeneous spherical cloud is given in Fig. 28.
Inside a cloud, the retarded cooling wave, which is associated
with a high peak at the curve of cooling in the area of hydrogen–
helium recombination, is formed. An inverse temperature dependence
of the cooling rate results in the development of a thermal
instability and the appearance of a thick cold layer between the
hot center and thin external heated layer. It is interesting to
note that the phase velocity of the external boundary of the
cooling wave exceeds the velocity of light, which leads to the
con- finement of photons produced in the recombination inside the
cold layer [25, 10].

\begin{figure}
\centerline {\psfig{file=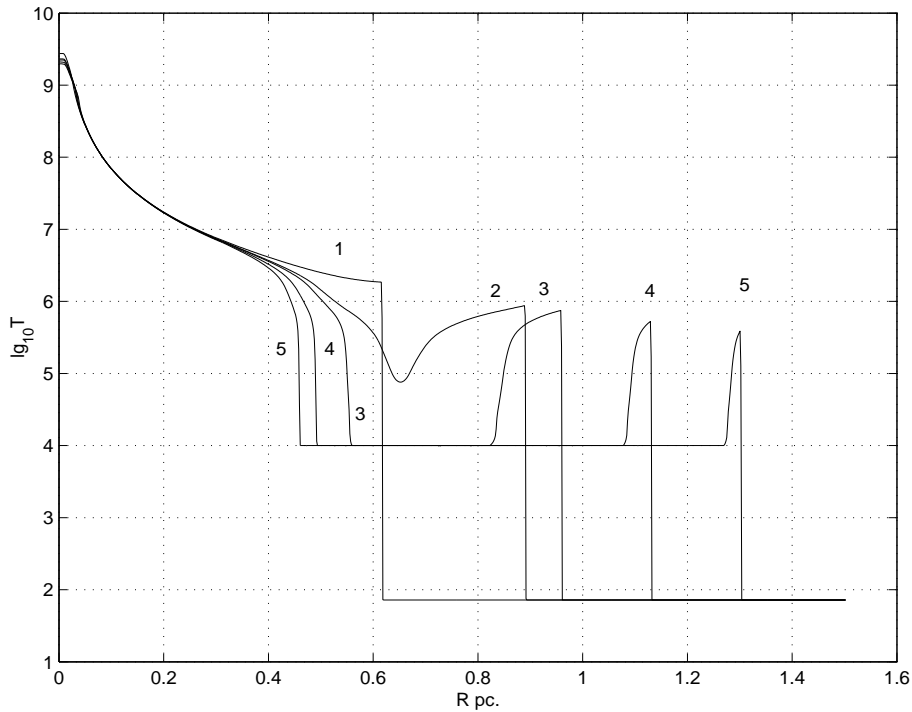,width=12cm,angle=-0}
 }
\caption{Temperature distribution in a cold homogeneous molecular
cloud with the radius $R = 1.5$ pc, the hydrogen concentration
$n_H=10^5$~cm$^{-3}$, the normal chemical composition after a
homogeneous GRB flare of the total energy $Q=5\cdot10^{52}$ erg,
with the flat spectrum for $E_{max}=2$ MeV at the consecutive
moments of time after the GRB: (1) 2.02, (2) 2.91, (3) 3.14, (4)
3.70, and (5) 4.26 yr [10].} \label{Figure_40}
\end{figure}
The molecular cloud can be nonuniform, and the GRB flare can take
place generally outside its center. It is conventionally assumed
that the GRB radiation should be anisotropic being a narrow beam.
The results of calculations of the distribution of density for
such a general case are shown in Fig. 29. Here, the cooling
proceeds faster in the higher-density region so that a thin hot
layer is formed behind the GRB-propagation front, and a much
thicker hot layer remains in the lower-density region behind the
GRB.

\begin{figure}
\centerline
{\psfig{file=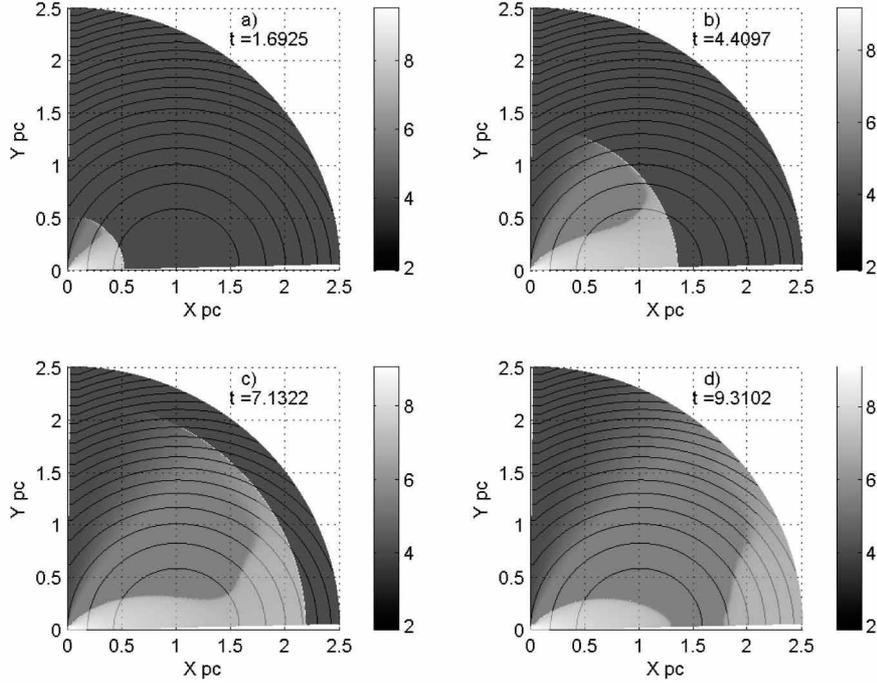,width=12cm,angle=-0}
 }
\caption{Temperature distribution (dark filling, $\log{T}$ is
indicated in the column on the right) and density levels (lines)
after the GRB in the cold molecular cloud of a radius $R = 1.5$ pc
with a hydrogen concentration
$n_H=10^5e^{\displaystyle{-r^2/r_0^2}}$cm$^{-3}$, $r_0=0,75$ pc,
with the normal chemical composition after an anisotropic GRB with
the angular distribution of radiated energy in the form
$Q=5\cdot10^{54}e^{\displaystyle{ -\left(\frac{\theta'}
{\theta_0}\right)^2}}$, $\theta_0=0,4$ rad, and a flat spectrum
with $E_{max}=2$ MeV located at the distance $x_0=1,0$ pc from the
molecular-cloud center. All distances are indicated in parsecs and
the time in years [10].} \label{Figure_41}
\end{figure}

\subsection{Hydrodynamic motions of gas caused by cosmological GRB}

The maximal velocities attained by the matter of a homogeneous
molecular cloud after the GRB flare inside it with a total energy
release of $1.6\cdot10^{53}$ or $1.6\cdot10^{54}$ erg and the
maximal photon energy $E_{max}=2$ MeV in a flat spectrum are equal
to $2,16 \cdot 10^3$ and $5,17 \cdot 10^3$ km/s, respectively. The
further increase in the GRB energy does not lead to increasing the
velocity behind the shock-wave front because of the
Compton-heating properties [10]. Fast cooling of matter hampers
shockwave enhancement even in the case of a nonuniform cloud with
a moderate density gradient.

It is possible to achieve a higher acceleration of matter after
the GRB flare in some cases of a special distribution of matter
around the GRB. When the GRB flares inside a low-density cone
inside the molecular cloud, the density gradient is formed inside
this cone and results in the jet of matter with a high velocity
along the cone axis. The molecular-cloud matter nearby the GRB
including that in the cone is heated to almost identical
temperatures so that the arising pressure gradient follows the
initial density gradient. The motion of matter to the cone axis
under the action of this gradient results in the appearance of a
cumulative jet along the cone axis, the matter in which is
accelerated to the velocity of $1,7\cdot 10^4$ km/s. The evolution
of the temperature distribution and the induced velocity field are
shown in Figs. 30 and 31. The conic-type density distribution
arises when the GRB flares between molecular clouds in the
low-density region (in this case, the jet is ringshaped), or the
similar cone can be formed by an anisotropic star wind of a
GRB-progenitor star.

\begin{figure}
\centerline {\psfig{file=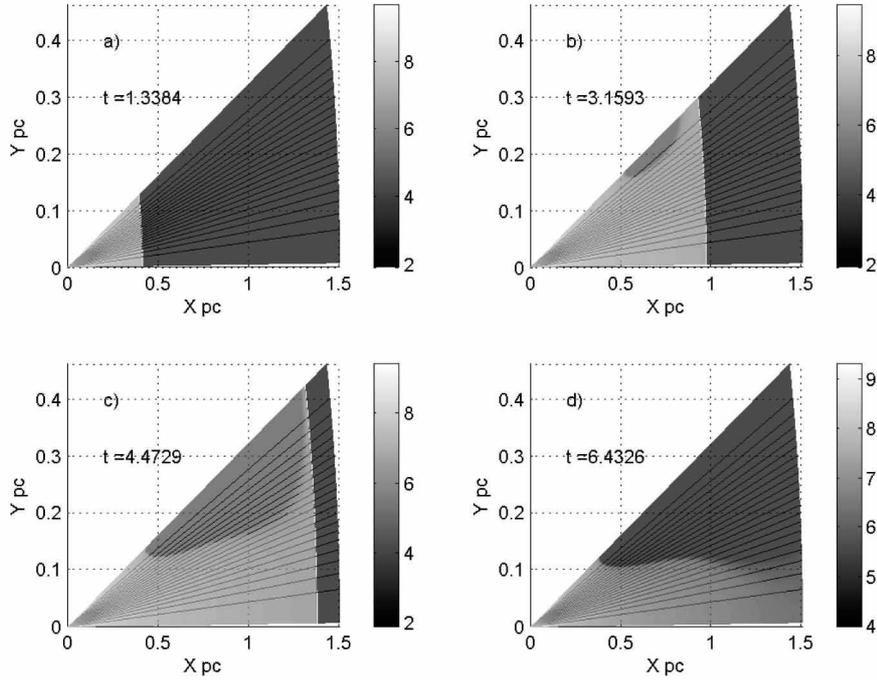,width=12cm,angle=-0}
 }
\caption{Distributions of velocity (arrows) and temperature (dark
filling, $\log{T}$ are indicated in the column on the right) in
the initially cold molecular cloud of the radius $R = 1.5$ pc at
various moments of time after an isotropic GRB for the variant
with an initial hydrogen- concentration distribution of the conic
type $n_H=10^5 e^{{-2-2\cos(10\theta)}}$ cm$^{-3}$. The calculated
region is limited to the angles $0 \leq \theta \leq \pi/10$, while
$v_{\theta}=0$, on the boundary $ \theta = \pi/10$; the distance
from a GRB source is indicated in pc, and the time—in yr [10].}
 \label{Figure_43}
\end{figure}

\begin{figure}
\centerline {\psfig{file=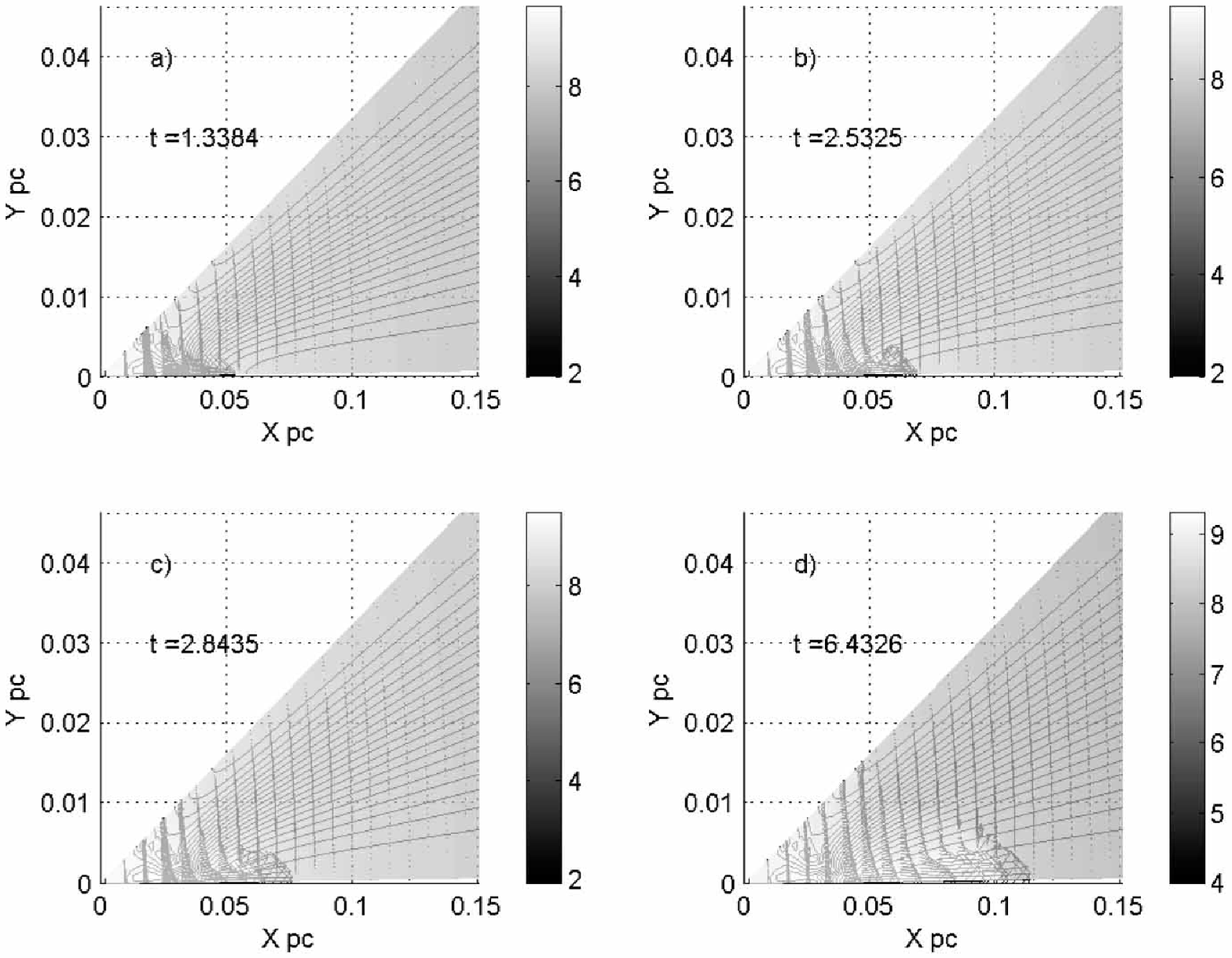,width=12cm,angle=-0}
 }
 \caption{Zoomed central part of Fig. 30 [10].}
 \label{Figure_44}
\end{figure}
The afterglows from an anisotropic GRB flare can have a strong
angular dependence because of a strong absorption of light by
dust, which is not evaporated after the flare along the directions
perpendicular to the GRB axis. This fact can weaken the
restrictions on a collimation angle following from the lack of
optical orphan bursts because the optical afterglow of similar
type is collimated almost the same as the GRB itself. The
reradiation on the dust leads to the appearance of an infrared GRB
afterglow, which should be more isotropic than the optical one.
Therefore, the search for infrared orphan bursts accompanying an
anisotropic GRB seems to be more promising.

 \section{Afterglows in high-energy region}

Observations in the EGRET experiment onboard of the orbital
Compton г-ray observatory (CGRO) show that the GRBs radiate also
in hard gamma regions up to an energy of 20 GeV [35]. Hard
$\gamma$ rays were detected approximately for 10 GRBs, photons
with energies over 100 MeV being observed for 5 GRBs listed in
Table 3 from [93]. As a rule, the hard $\gamma$ rays last longer
than the basic soft GRB by up to 1.5 h in GRB 940217 (see Fig.
32). The comparison of the angular aperture in the experiments
EGRET and BATSE and also the duration of their operation at the
orbit results in the conclusion that the hard $\gamma$ rays should
be observed during a significant part (approximately one third to
half) of the entire GRB. The slope of the spectrum in the hard
gamma region is in the range between –2 and –3.7 and quickly
varies becoming softer with time (GRB 920622 from [94]). According
to the observations in the hard gamma region of a radio pulsar in
the Crab nebula [73] and PSR B1055-52 [102], the slope and the
variation of their spectra have properties similar to those of a
GRB. Taking into account the nonpulsing component of the
Crab-nebula radiation, the slope of its gamma spectrum varies
between –1.78 and –2.75. If a GRB is associated with the supernova
explosion and the neutron-star birth, the residual neutron-star
vibrations after explosion can be responsible for a long afterglow
in the hard gamma region [16, 103, 33].

\begin{table}
\caption{Observations of powerful GRBs in the experiment EGRET
from [93]}
%\begin{center}
\medskip
\begin{tabular}{ccccc}
\hline\noalign{\smallskip} {GRB name}& {Maximal energy }&
{Radiation}& {Spectral}&
{Delayed}\\
& {(GeV)}& {duration}& {function}&
{radiation}\\
\noalign{\smallskip}
\hline\\
 GRB910503 & $10$ & $84$ s & $E^{-2.2}$ & X \\
 GRB910601 & $0.314$ & $200$ s & $E^{-3.7}$ & X \\
 GRB930131 & $1.2$ & $100$ s & $E^{-2.0}$ & X \\
 GRB940217 & $18$ & $1.5$ h & $E^{-2.6}$ & X \\
 GRB940301 & $0.16$ & $30$ s & $E^{-2.5}$ &  \\
\hline\\
\end{tabular}
%\end{center}
\label{tab4}
\end{table}

\begin{figure}
\centerline {\psfig{file=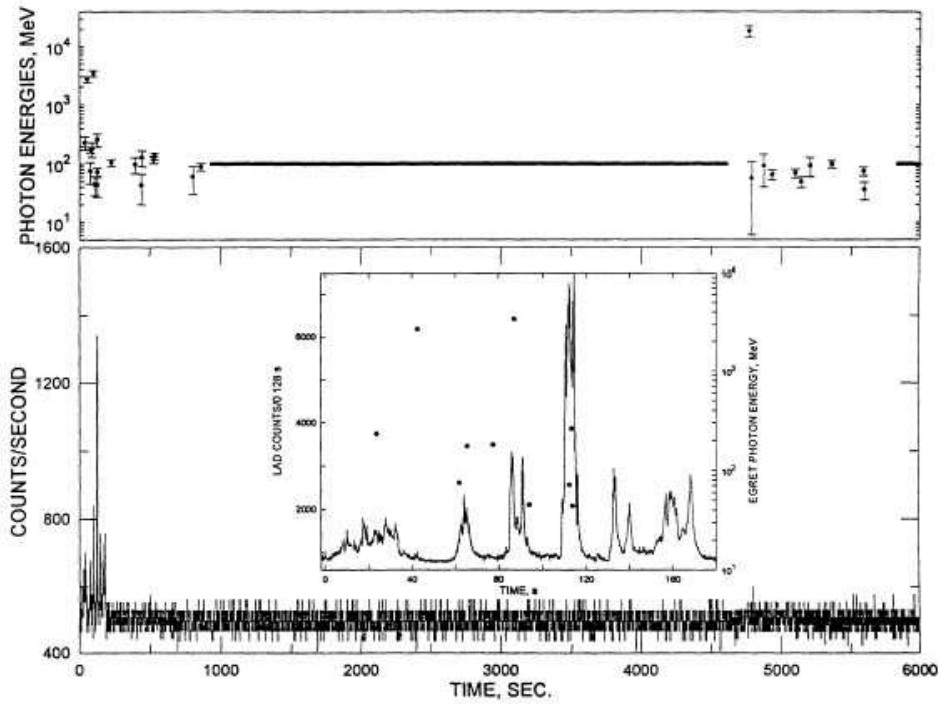,width=13cm,angle=-0}
 }
\caption{GRB flare of February 17, 1994, the gamma rays from which
were observed in GeV-energy region in the experiment EGRET within
1.5 h after the basic flare. The composite figure includes the
EGRET data and the data of observations of the basic flare in the
BATSE experiment and at the satellite Ulysses [35].}
\label{grbfig9}
\end{figure}
All existing projects in the hard gamma region assume the
investigation of GRBs as the primary problem. For this purpose, it
is required to have the angular aperture as large as possible. It
is expected that the nearest-in-time launch will be that of the
small astronomical gamma satellite of the Italian Cosmic Agency
(AGILE) of about 130 kg in weight, which should operate within 3
years in the energy range from 30 MeV to 30 GeV according to the
project. In the AGILE experiment, it is proposed to use
solid-state silicon detectors the fabrication technology of which
is well developed. It is proposed to achieve the optimum angular
resolution when constructing the images in the gamma region at a
level of $5^\prime$ –- $20^\prime$ for powerful sources and to
perform the observation in an unprecedented large field of view of
about 3 sr. This results in an effective sensitivity which is
comparable with that of the EGRET for point sources observed along
its axis and much higher for the sources away from the axis. The
detection of GRB is expected in the range above 50 MeV at a level
of 5 –- 10 events per year [1].

The gamma large-area space telescope (GLAST), slated for launch
later, is an international space project with the participation of
the space agencies of various countries. The project is purposed
for the space investigations in the energy range from 10 keV to
300 GeV. After the realization of several successful research
space projects in the field of gamma-ray astronomy, the EGRET
project was realized onboard of the CGRO launched in 1991. In the
EGRET experiment, the first complete survey of the sky in the
energy range from 30 MeV to 10 GeV was performed, and the majority
of sources discovered in this survey remained unidentified. In the
GLAST experiment, the investigations of the gamma region will be
carried out in the range 20 MeV –- 300 GeV. The comparative
characteristics of GLAST (EGRET) are as follows. The maximal
effective region amounts to 8000 cm$^2$ (1500 cm$^2$), the field
of view exceeds 2 sr (0.5 sr), the angular resolution is better
than 3.5$^\circ$ at 100 MeV and better than 0.15$^\circ$ at
energies higher than 10 GeV (5.8$^\circ$ at 100 MeV), the energy
resolution is better than 10\% (10\%), and the localization
accuracy (radius at the  1-$\sigma$ level) for steady
high-altitude sources with large $\vert b \vert$ and a flux of
10$^{-7}$ cm$^{-2}$s$^{-1}$ in the energy range higher than 100
MeV amounts to less than $0.5^\prime$ (15$^\prime$). On board
GLAST, the almost-all-sky monitor in the soft gamma region is
specially mounted to obtain simultaneously the data in soft and
hard gamma-photon energy regions for the independent detection of
GRBs. As in the experiment AGILE, the detector for detecting hard
gamma photons comprises 18 silicon strips. This technology has a
long and successful history of application in experimental
high-energy physics in accelerators. It has a high detection
efficiency ($>$99\%), an excellent positional resolution ($<$60
$\mu$m in this design), and a high signal-to-noise ratio ($>$20 :
1). GLAST is expected to have high detection efficiency at
energies above 10 GeV and the possibility of GRB localization with
accuracy reasonably high for the fast search of their radiation at
all larger wavelengths. GLAST is expected to detect about 200 GRBs
per year, half of them located in the radius of no more than 10
arcmin, giving a good possibility to obtain the image of the
region with GRBs using a large-aperture optical telescope [72].

The original design of the gamma telescope in the energy region
higher than 0.5 -- 1 GeV was considered in [24, 58, 31] and named
CYGAM—Cylindrical Gamma Monitor. The principal schematic drawing
of CYGAM is shown in Fig. 33. The hollow inside cylindrical shape
allows the authors to reduce the weight of the device several
times for the same sensitivity and the best angular resolution at
an energy of about 1 GeV. In this scheme, the drift chambers [31]
were stipulated for detectors; however, the above silicon strips
can also be used instead of them. It is possible to expect that,
in the energy range of 1 -- 10 GeV, the CYGAM should have
characteristics comparable to or even exceeding those of the
GLAST, being almost three times lighter and, correspondingly,
cheaper to manufacture. The large field of view of approximately 7
sr will be highly suitable for detecting GRBs.

\begin{figure}
\centerline {\psfig{file=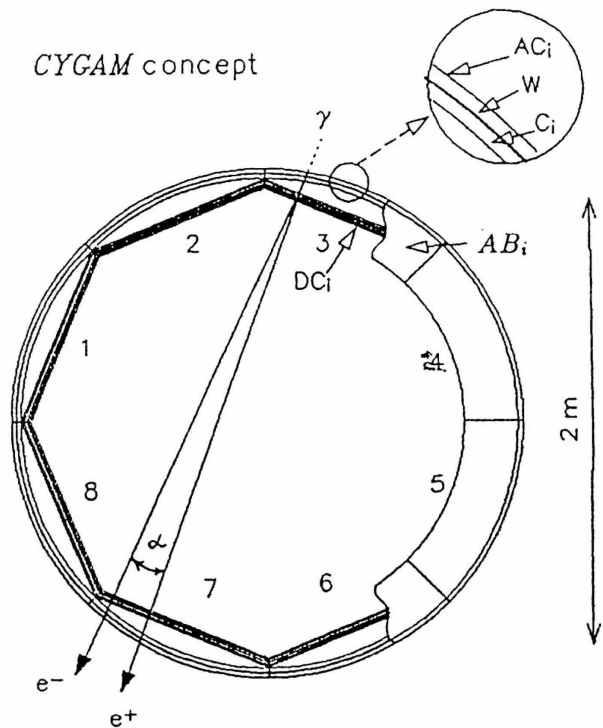,width=9cm,angle=-0}
 }
 \caption{Schematic diagram of proposed CYlindrical
GAmma Monitor (CYGAM). $ AC_i $ are the
coincidence–anticoincidence counters, $ AB_i $ is the butt-end
anticoincidence shield, $W$ is the lumped converter, $ C_i $ are
the coincidence counters; and $ DC_i $ are the drift-chamber
layers [24].}
 \label{cygam}
\end{figure}

\section{Lines in the hard $x$-ray region}

Lines in the hard x-ray region of GRB spectra were discovered in
the experiment KONUS [64]. They were interpreted as cyclotron
lines and observed in approximately 30\% of GRBs. The lines and
spectra were strongly variable: an observed absorption-line depth
decreased in time (Fig. 34). The BATSE detector had the lower
spectral resolution, and, for a very long time, no lines were
found in the observed spectra. In later publications, at last, the
presence of spectral details in the hard-x-ray region of BATSE GRB
spectra was reported [29]. In this paper, when investigating the
spectra of 117 GRBs, statistically significant lines were found in
13 GRB spectra. A good example is GRB941017 for which the
consistent lines are well seen in the spectra of two detectors.
For some GRBs, one detector observed a line, while it proved to be
statistically insignificant for another detector. Figure 35 shows
the spectrum of GRB 930916 with the statistically significant line
from [29]. In this paper, the authors affirmed that the doubts
about the reality of the spectral details of GRBs in the hard
$x$-ray region will remain until the causes for the differences
between spectra of the same GRB observed by different detectors
are understood. However, we note that the GRB 930916 spectra were
obtained only within 20 s after the flare onset, and, according to
[64], the strongest spectral details were observed at the GRB
initial stage (see Fig. 34). The only study devoted to the
interpretation of such lines in the cosmological model [40] is
based on considering the spectrum strongly shifted in the blue
region (the corresponding relativistic factor $\Gamma = 25 - 100$)
for the gas cloud illuminated by the fireball $\gamma$ rays. A
similar model was proposed in [21] for an explanation of the lines
observed in the experiment KONUS within the framework of the GRB
galactic model [23]. The model was based on the explosion near the
neutron-star surface resulting in the jet of an expanding cloud
with $v/c = 0.1 - 0.3$ in which a line absorption occurs (see Fig.
36). In this model, the observed attenuation in the absorption
versus time (Fig. 34) was related to a decrease of the absorption
thickness during the expansion of the cloud.

\begin{figure}
\centerline{ \psfig{file=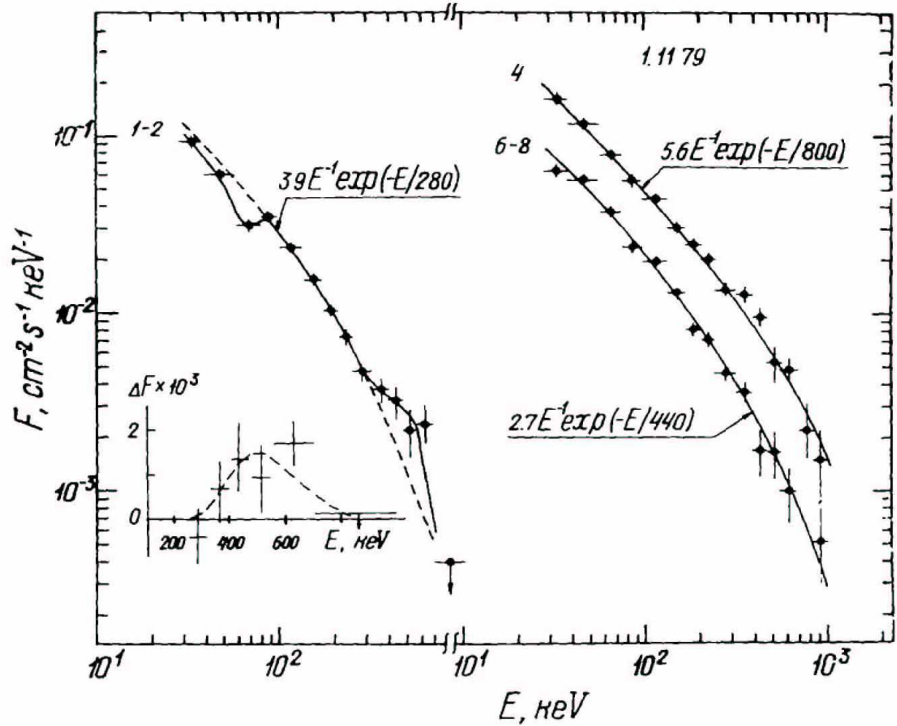,width=14cm,angle=-0} }
\caption{Spectral evolution of GRB of November 1, 1979. (1–2) The
spectrum obtained during the first 8 s with the absorption line at
an energy of $\approx 65$ keV and a wide emission structure in the
region of 350 -- 650 keV, (4) the spectrum of the fourth 4-s
interval, and (6–8) the total spectrum of the sixth, seventh, and
eighth intervals [65].} \label{fig8ma}
\end{figure}

\begin{figure}
\centerline {\psfig{file=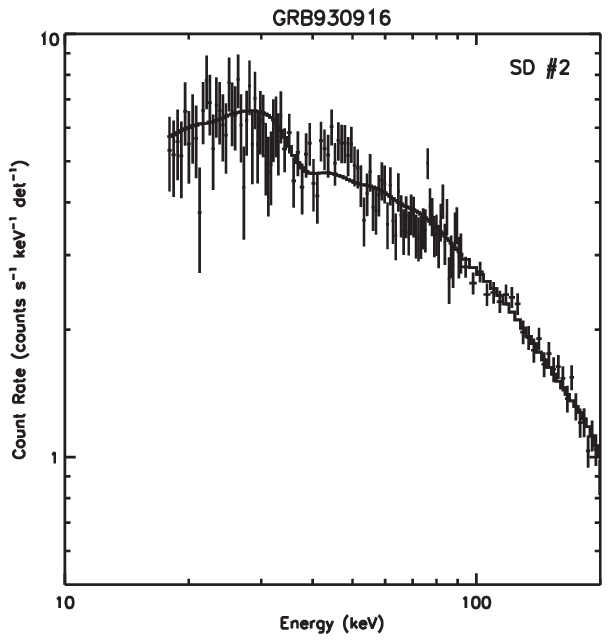,width=6.6cm,angle=-0}
 \psfig{file=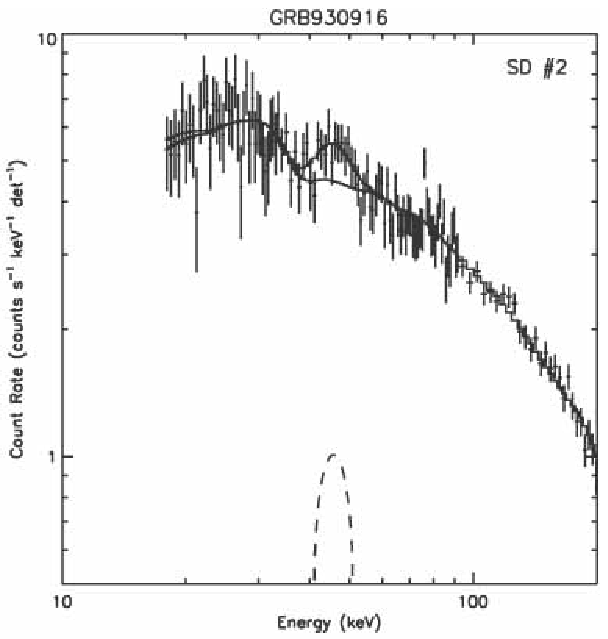,width=6.6cm,angle=-0}
 }
   \caption{Spectrum of GRB 930916 obtained from the BATSE
   observations in the interval from 22.144 to 83.200 s after its detection.
The bump at 30 keV is associated with the K-boundary of iodine in
the NaI detector. On the left: the best fit of data from the
detector SD~2 with the continuum spectrum. On the right: with the
addition of a narrow spectral line of radiation at 45 keV, the
$\chi^2$ value is improved by 23.1. Solid line corresponds to the
total spectrum, and strokes separately give the continuum spectrum
and the line [29].}
   \label{batlines}
\end{figure}

\begin{figure}
\centerline {\psfig{file=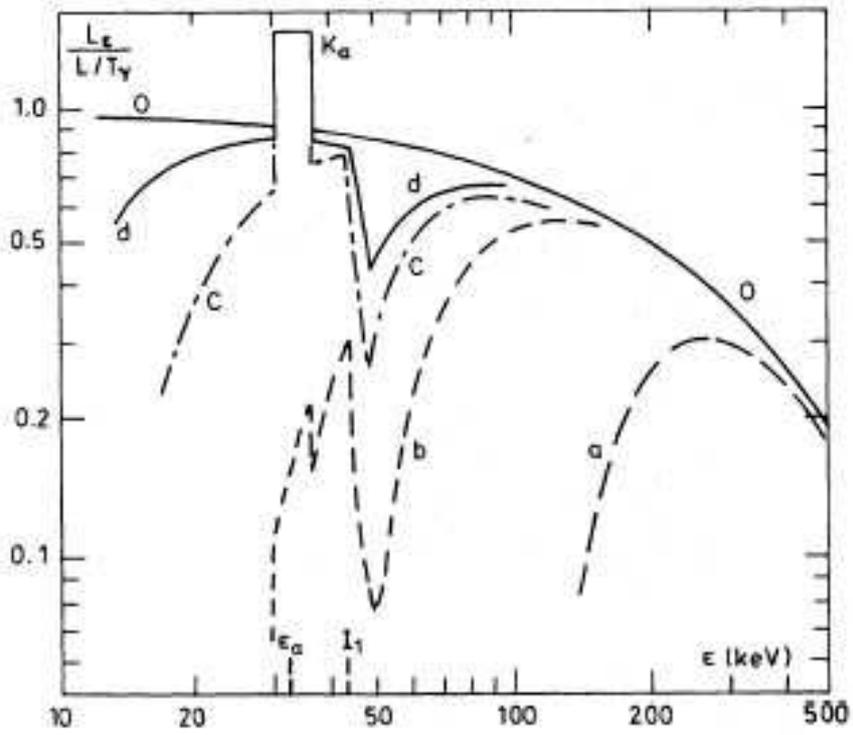,width=12cm,angle=-0}
 }
\caption{Time evolution of the absorption line formed due to the
presence of barium ions with $X_{Ba}=1/300$ in the iron plasma of
an expanding cloud located between a GRB source and an observer,
$t_a<t_b<t_c<t_d$; “0” is the GRB spectrum before the passage
through a cloud [21].} \label{bki}
\end{figure}

\section{GRB and origin of ultrahigh-energy
cosmic rays (UHECRS)}

In [69], it was noted that the squares of errors of two events
associated with cosmic rays (CRs) of the highest energies overlap
with the squares of errors of strong GRBs. GRB 910503 and GRB
920617 were observed 5.5 and 11 months before the corresponding
CRs, respectively. In one of the cases, the most powerful CRs
coincide with the most powerful GRB from the BATSE catalogue.
Accidental coincidences of such events is very low; therefore, it
was assumed in [69, 70] that the same phenomenon results in the
occurrence of both GRBs and ultrahigh-energy CRs. A time delay, as
well as a small divergence at the celestial sphere is expected
because the magnetic field affects the trajectory of a CR charged
particle and makes it curvilinear. The estimates show that only
the Galaxy field taken alone can lead to an observed delay of CRs
in comparison with GRBs. If we assume that the energy release in
CRs and GRBs is approximately identical [69], several tens of
percents of the strongest GRBs can explain the available
experiments on detecting ultrahigh-energy CRs (UHECRs).

The assumption that GRBs are the sources of UHECRs was almost
simultaneously made in [110, 111, 108]. The isotropy of
distribution of both events in the sky and their cosmological
origin point to the GRB– UHECR relation. In addition to this, the
observed energy-release rate (per unit volume) in GRB $\gamma$
rays proves to be comparable with the high-energy-proton energy,
which is necessary for explaining the observable UHECR flux. The
energy-release rate was estimated in [112] (see also [109]). It
was assumed that the extragalactic protons in the energy range
from $10^{19}$ to $10^{21}$ eV are produced in cosmologically
distributed sources with the following density and spectrum:

\begin{equation}
E_p^2\frac{d\dot{N}_p}{dE_p}\approx \,0.6\times 10^{44}\,
\mbox{erg}\, \mbox{Mpc}^{-3} \mbox{yr}^{-1}. \label{eq:energyrate}
\end{equation}
The energy-release rate in a GRB can be estimated assuming that a
local ($z = 0$) GRB frequency amounts to $\approx0.5 {\rm Gpc^{-3}
yr^{-1}}$ [92]. In this case, an average GRB energy with known
redshifts in the MeV range of the $\gamma$-photon energy is equal
to $\approx0.5/({\rm Gpc^3 yr})$ erg. The GRB frequency with known
redshifts is less than the total observed GRB frequency by a
factor of $\approx0.7\times0.5=0.35$. The factor 0.7 is associated
with the GRB detection threshold by the satellite Beppo-SAX, which
is approximately two times higher than that found from the BATSE
data used for estimating the total GRB frequency. The factor 0.5
is associated with the fact that optical afterglows are detected
only for approximately half of GRBs. Thus, the local ($z = 0$)
energy release rate in the MeV-photon region during the GRB flares
is equal to

\begin{equation}
\label{eq:grbrate}
    \dot{\varepsilon}_{\gamma [\mbox{MeV}]}\ge0.35\times 3\times10^{53}\mbox{erg}
    \times 0.5/\mbox{Gpc}^3/\mbox{yr}=0.5\times10^{44}\mbox{erg/Mpc}^3 /\mbox{yr}.
\end{equation}
This energy-release rate is comparable with that for UHECRs from
Eq. (11).

The energy-release rate can be estimated in another way using the
luminosity function obtained in [92]. On the basis of the large
set of GRB events, it was found that the number of GRBs as a
function of the peak luminosity $L_c$ follows a power law in the
range from $\sim 10^{50.5}$ to $\sim 10^{51.5}$ erg/s and then
abruptly falls. Using a value of $\sim~10^{51}$ erg/s for the peak
luminosity and 10 s as an average equivalent time of the total
energy release, we estimate the total GRB energy-release rate to
be approximately an order of magnitude less than that followed
from Eq. (12). This value is more than ten times less than the
UHECR energy-release rate from Eq. (11). With taking into account
the large number of uncertainties, even the divergence in order of
magnitude cannot be the basis for the final refusal to consider
the hypothesis about the GRB - UHECR relation.

\section{Short GRBs and SGRs}

The cosmological GRB origin creates many problems when
constructing a realistic physical model. The basic problem
consists in the difficulty of obtaining an enormous energy release
over such a short time in a small volume. For a collimated GRB,
the requirements on the energy release decrease; however, the
collimation angle cannot be too small. Observations of optical
orphan bursts during the complete sky survey by optics can
essentially improve the restrictions on the collimation angle. The
measurements of optical spectra of prompt afterglows, while
optical luminosity is still reasonably high, and the investigation
of the polarization in optical and x-ray regions are very
important at this stage for clarifying the radiation mechanism.

It is probable that GRBs do not represent a homogeneous set of
events and consist in bursts of completely different origins. The
statistical analysis reveals at least two separate sets consisting
of long-duration ($>$2 s) and short-duration bursts. We note that
optical afterglows with determining the redshifts were measured
only for long bursts. Therefore, it is possible that the short
bursts are of another origin (possibly, even galactic).

It is interesting to compare the properties of short GRBs with
giant flares of soft gamma repeaters (SGRs) located inside the
Galaxy. If the SGRs were located at a greater distance on which a
usual activity of SGRs would not be seen, and only giant bursts
would be detected, these last would be attributed, without doubt,
to conventional short GRBs. The light curves of two giant bursts
are shown in Figs. 37 and 38 according to the observations by the
KONUS–-WIND device [66, 67]. The frequency and power of giant SGR
bursts observed already in all four reliably identified SGRs in
the Galaxy and in the Large Magellanic Cloud (LMC) are such that
there is a real possibility of their observations from local-group
adjacent galaxies as short GRBs. The estimates show that, by
statistics, more than 10 “short GRBs” of similar type should be
observed from the Andromeda Nebula (M 31) and other local-group
galaxies [18]. No GRBs in nearby adjacent galaxies could point to
the fact that the observed SGRs are closer and weaker objects than
it is believed now on the basis of their possible genetic relation
to supernova remnants (SRs), the distances to which can be
estimated. It is possible that this relation does not exist at all
[18].

\begin{figure}
\centerline {\psfig{file=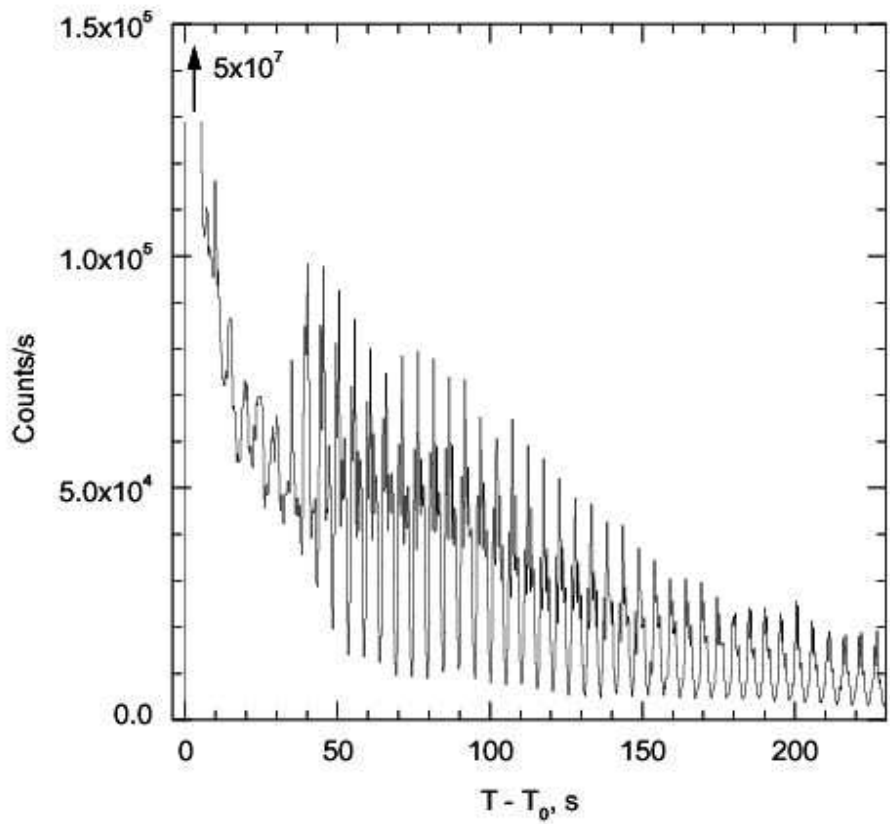,width=13cm,angle=-0}
 }
\caption{Light curve of the giant burst SGR 1900 + 14 of August
27, 1998. The presented data on the radiation in the photon-energy
region $E$ $>$ 15 keV are from [66].} \label{sgrm1}
\end{figure}

\begin{figure}
\centerline {\psfig{file=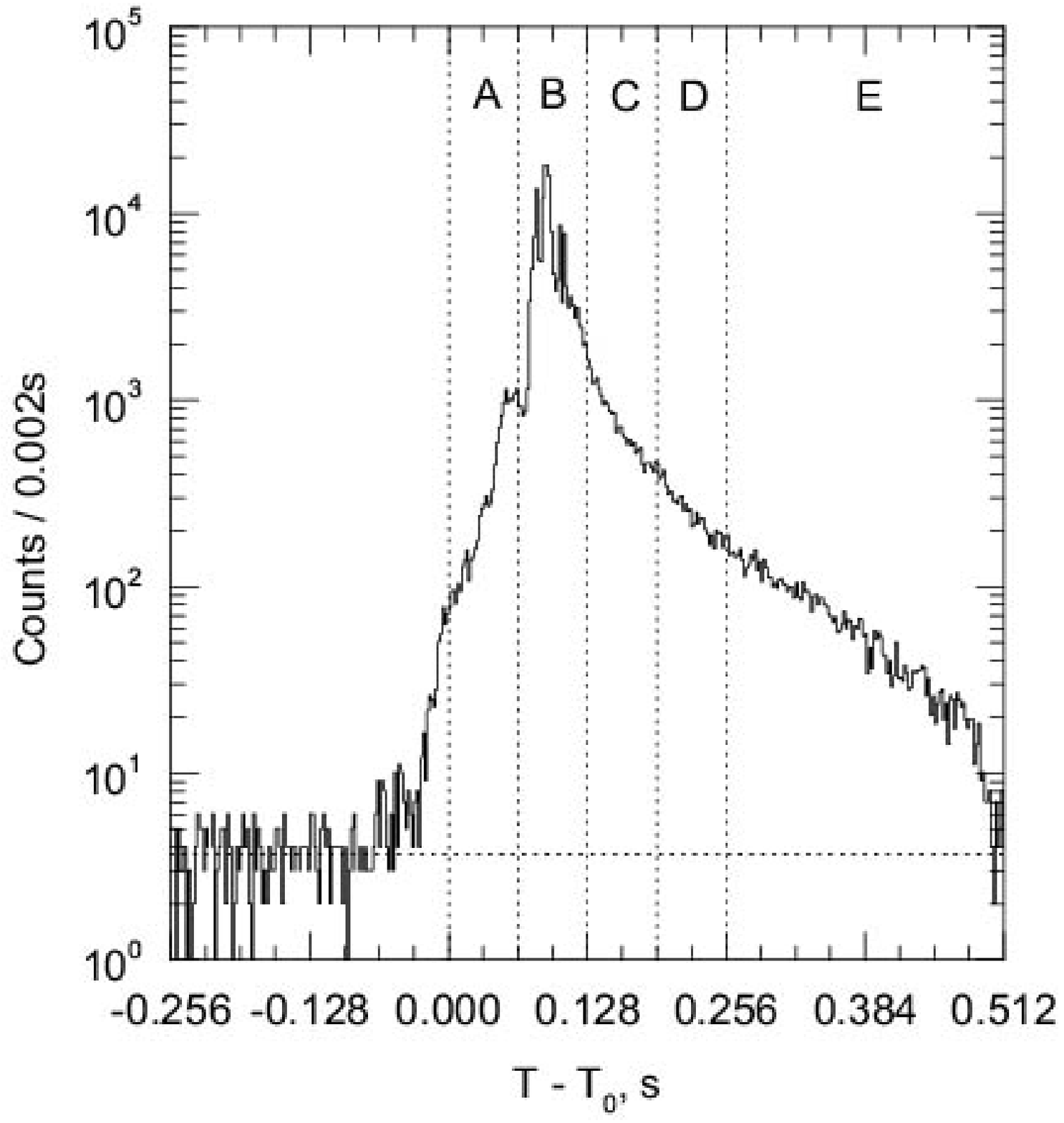,width=13cm,angle=-0}
 }
\caption{Light curve of the giant burst SGR 1627-41 of June 18,
1998. The presented data correspond to the radiation in photon
energy region $E$ $>$ 15 keV. Time of rise of brightness amounts
approximately to 100 ms [67].}. \label{sgrm2}
\end{figure}

\section{Conclusions}

Despite more than 30 years of observations and numerous attempts
at constructing theoretical models, until now, there is no
authentic model either for the formation of observable radiation
or for the GRB energy source. Although long GRBs are most likely
of the cosmological origin, it is possible that short GRBs have
another origin (probably, even galactic) and are associated with
giant SGR flares [65].

Critical experiments, which could finally clarify the GRB nature,
are associated with observations in all regions of the
electromagnetic spectrum. In optics (from infrared to
ultraviolet), it is necessary to obtain GRB spectra at the
earliest stages as close as possible to the time of the г-ray
burst itself. This problem is formulated for wide-angle automatic
telescopes, which already operate or are being designed in various
countries. They should be quickly directed to a sky region using
the information obtained from a gamma satellite right after the
detection of a GRB. Similar problems is solved by the specialized
satellite SWIFT launched on November 20, 2004 and purposed for
observations of GRBs in a wide range from hard x rays (150 keV) to
the optical band.  SWIFT registers about 100 GRBs with a
localization accuracy of about 1' within one year of operation. An
important observational problem in optics is the search for the
orphan bursts not accompanied by a $\gamma$-ray pulse. about the
degree of collimation of the gamma pulse. Such observations
require a long-term all-sky scanning using optics; they are
possible with either a system of continuously operating very
wide-angle telescopes covering the entire sky or a specialized
satellite with an optical telescope (or system of telescopes)—the
all-sky monitor. In addition to the gamma observations for
detecting GRBs in the energy region of about one MeV, the
essential information can be obtained from observations in the
hard gamma region with a photon energy ranging from a hundred MeV
to tens of GeV. Such observations in the same energy range are
planned to be carried out by the gamma large-angle space telescope
(GLAST) with the expected launch in 2007 and by the gamma small
space telescope AGILE, which should be launched earlier. The
collapse of a high-mass star nucleus with the formation of a black
hole surrounded by a massive rotating disk represents the probable
cosmological GRB model. The fast fall of the disk onto the black
hole as a result of the magnetorotational processes accompanied by
an enormous energy release in the form of neutrino, radiation, and
electron–positron pairs can be the origin of GRBs. The possibility
of originating GRBs in exotic processes associated with cosmic
strings, antimatter, etc., is also not excluded.

\end{document}